\documentclass[twocolumn]{aastex63}
\usepackage[utf8]{inputenc}

\usepackage{graphicx}	
\usepackage{amsmath}	
\usepackage{amssymb}	
\usepackage{xspace}	    

\newcommand{\simgt}{\,\rlap{\lower 3.5 pt \hbox{$\mathchar \sim$}} \raise 1pt \hbox {$>$}\,}
\newcommand{\simlt}{\,\rlap{\lower 3.5 pt \hbox{$\mathchar \sim$}} \raise 1pt \hbox {$<$}\,}
\newcommand{\dd}{\mathrm{d}}

\newcommand{\BE}{\begin{equation}}
\newcommand{\EE}{\end{equation}}
\newcommand{\BEA}{\begin{eqnarray}}
\newcommand{\EEA}{\end{eqnarray}}

\newcommand{\DV}{\ifmmode{\Delta v}\else $\Delta v$\xspace\fi}

\newcommand{\HI}{\ifmmode{\textsc{hi}}\else H\textsc{i}\fi\xspace}
\newcommand{\HII}{\ifmmode{\textsc{hii}}\else H\textsc{ii}\fi\xspace}

\newcommand{\MUV}{\ifmmode{M_\textsc{uv}}\else $M_\textsc{uv}$\xspace\fi}
\newcommand{\fesc}{\ifmmode{f_\textrm{esc}}\else $f_\textrm{esc}$\xspace\fi}
\newcommand{\lya}{\ifmmode{\mathrm{Ly}\alpha}\else Ly$\alpha$\xspace\fi}

\newcommand{\nh}[1][]{\ifmmode{\overline{n}_\textsc{h}^{#1}}\else $\overline{n}_\textsc{h}$\xspace\fi}

\newcommand{\xHI}{\ifmmode{x_\HI}\else $x_\HI$\xspace\fi}
\newcommand{\xHImean}{\ifmmode{\overline{x}_\HI}\else $\overline{x}_\HI$\xspace\fi}
\newcommand{\xHIImean}{\ifmmode{\overline{x}_\HII}\else $\overline{x}_\HII$\xspace\fi}
\newcommand{\trec}{\ifmmode{t_\textrm{rec}}\else $t_\textrm{rec}$\xspace\fi}
\newcommand{\clump}[1][]{\ifmmode{C_\HII^{#1}}\else $C_\HII$\xspace\fi}

\newcommand{\Nion}{\ifmmode{\dot{N}_{\mathrm{ion}}}\else $\dot{N}_\mathrm{ion}$\xspace\fi}
\newcommand{\Rion}[1][]{\ifmmode{R_\mathrm{ion}^{#1}} \else $R_\mathrm{ion}$\xspace\fi}

\newcommand{\kms}{\,\ifmmode{\mathrm{km}\,\mathrm{s}^{-1}}\else km\,s${}^{-1}$\fi\xspace}
\newcommand{\cm}{\,\ifmmode{\mathrm{cm}}\else cm\fi\xspace}

\begin{document}
\title{The Evolution of the Lyman-Alpha Luminosity Function During Reionization}

\accepted{July 1, 2021}
\submitjournal{The Astrophysical Journal}

\correspondingauthor{Alexa M.\ Morales}
\email{amora250@fiu.edu}


\author[0000-0003-4965-0402]{Alexa M.\ Morales}
\affil{Florida International University, 11200 SW 8 St., Miami, FL 33199, USA}
\affiliation{Center for Astrophysics $|$ Harvard \& Smithsonian, 60 Garden St., Cambridge, MA 02138, USA}

\author[0000-0002-3407-1785
]{Charlotte A.\ Mason}\altaffiliation{Hubble Fellow}
\affiliation{Center for Astrophysics $|$ Harvard \& Smithsonian, 
60 Garden St., Cambridge, MA 02138, USA}

\author[0000-0000-0000-0000
]{Sean \ Bruton}
\affiliation{Minnesota Institute for Astrophysics, University of Minnesota, 116 Church St SE, Minneapolis, MN 55455, USA}

\author[0000-0003-2491-060X
]{Max \ Gronke}\altaffiliation{Hubble Fellow}
\affiliation{Department of Physics \& Astronomy, Johns Hopkins University, Baltimore, MD 21218, USA}

\author[0000-0003-3291-3704
]{Francesco \ Haardt}
\affiliation{DiSAT, Università dell’Insubria, via Valleggio 11, 22100 Como, Italy}
\affiliation{National Institute of Nuclear Physics INFN, Milano - Bicocca, Piazza della Scienza 3, 20126 Milano, Italy}

\author[0000-0002-9136-8876
]{Claudia \ Scarlata}
\affiliation{Minnesota Institute for Astrophysics, University of Minnesota, 116 Church St SE, Minneapolis, MN 55455, USA}

\begin{abstract}
The time frame in which hydrogen reionization occurred is highly uncertain, but can be constrained by observations of Lyman-alpha (Ly$\alpha$) emission from distant sources. Neutral hydrogen in the intergalactic medium (IGM) attenuates Ly$\alpha$~photons emitted by galaxies. As reionization progressed the IGM opacity decreased, increasing Ly$\alpha$~visibility. The galaxy Ly$\alpha$~luminosity function (LF) is thus a useful tool to constrain the timeline of reionization. In this work, we model the Ly$\alpha$~LF as a function of redshift, $z=5-10$, and average IGM neutral hydrogen fraction, $\overline{x}_\textsc{hi}$. We combine the Ly$\alpha$~luminosity probability distribution obtained from inhomogeneous reionization simulations with a model for the UV LF to model the Ly$\alpha$~LF. As the neutral fraction increases, the average number density of Ly$\alpha$~emitting galaxies decreases, and are less luminous, though for $\overline{x}_\textsc{hi} \lesssim 0.4$ there is only a small decrease of the Ly$\alpha$~LF. We use our model to infer the IGM neutral fraction at $z=6.6, 7.0, 7.3$ from observed Ly$\alpha$~LFs. We conclude that there is a significant increase in the neutral fraction with increasing redshift: $\overline{x}_\textsc{hi}(z=6.6)=0.08^{+ 0.08}_{- 0.05}, \, \overline{x}_\textsc{hi}(z=7.0)=0.28 \pm 0.05$ and $\overline{x}_\textsc{hi}(z=7.3)=0.83^{+ 0.06}_{- 0.07}$. We predict trends in the Ly$\alpha$~luminosity density and Schechter parameters as a function of redshift and the neutral fraction. We find that the Ly$\alpha$~luminosity density decreases as the universe becomes more neutral. Furthermore, as the neutral fraction increases, the faint-end slope of the Ly$\alpha$~LF steepens, and the characteristic Ly$\alpha$~luminosity shifts to lower values, concluding that the evolving shape of the Ly$\alpha$~LF -- not just its integral -- is an important tool to study reionization.
\end{abstract}

\section{Introduction}\label{sec:intro}

After Recombination, $\sim75\%$ of the baryons in the early universe were atomic hydrogen. In the present-day universe, the majority of hydrogen in the intergalactic medium (IGM) is ionized. At some point within the first billion years, ionizing photons, likely emitted by the first stars and galaxies, reionized hydrogen during this `Epoch of Reionization', initially in bubbles around galaxies which eventually overlapped and created an entirely ionized IGM \citep[e.g.,][]{Barkana2007,Mesinger2016,Dayal2018}.

The time frame in which reionization occurred is still highly uncertain. Its onset and progression are rather poorly constrained \citep[e.g.,][]{Greig2017,Mason2019b}. The best constraints currently come from observations of the increasing optical depth to \lya\ photons, observed in the spectra of high redshift quasars \citep[e.g.,][]{Fan2006,McGreer2015,Banados2018,Davies2018,Greig2017} and galaxies -- both those selected as Lyman-break galaxies \citep[e.g.,][]{Treu2013,Schenker2014,Mesinger2015,Mason2018,Mason2019,Hoag2019,Whitler2020,Jung2020} and Lyman-alpha emitters \citep[e.g.,][]{Malhotra2004,Konno2018}. These constraints imply a fairly late and rapid reionization \citep[e.g.,][]{Mason2019b,Naidu2020}, though c.f. \citet{Finkelstein2019,Jung2020} who find evidence for a slightly earlier reionization. 
During reionization, \lya photons are attenuated extremely effectively by neutral hydrogen \citep[e.g.,][]{Miralda1998,Mesinger2015,Mason2018}. As a result, \lya observations can be an investigative tool of the neutral IGM during the Epoch of Reionization. However, these reionization inferences are limited by systematic uncertainties in modelling the intrinsic \lya emission -- more independent probes are necessary to understand the systematic uncertainties in reionization inferences.

In this paper, we use the Lyman-alpha (\lya) luminosity function to constrain the progression of reionization with cosmic time.
\lya luminosity functions (LFs) have been used for over a decade to understand reionization \citep[e.g.,][]{Malhotra2001,Malhotra2004, Stern2005,Jensen2013}. LFs describe the luminosity distribution of a population of objects and we can quantify their evolution by looking at the LF at different redshifts. As \lya is typically expected to be the strongest emission line in the rest-frame optical to UV \citep[e.g.,][]{Peebles1967,Shapley2003}, wide-area ground-based narrow-band surveys \citep[e.g.,][]{Malhotra2004,Ota2008,Ota2010,Ouchi2010,Konno2014,Konno2016,Ota2017, Konno2018} and, more recently, space-based grism observations \citep[e.g.,][]{Tilvi2016,Bagley2017,Larson2018} have been efficient at discovering large populations of galaxies at high redshifts, selected based on strong \lya fluxes, and known as `\lya emitters' or LAEs. 

As \lya photons are obscured during reionization, a decline in the \lya LF is a signature of an increasingly neutral IGM. However, any evolution must be disentangled from the evolution in the underlying galaxy population with redshift (i.e., as galaxies become less numerous at high redshifts due to hierarchical structure formation). Previous works typically compared the evolution of the \lya LF to that of the UV LF, which describes the number density of Lyman-break galaxies and is not distorted by reionization, to establish the evolution due to neutral gas \citep[e.g.,][]{Ouchi2008,Ouchi2010, Konno2018, Konno2016}. These works estimated the neutral fraction at specific redshifts by using the drop in the \lya luminosity density compared to the UV luminosity density to calculate a transmission fraction, $T_\mathrm{IGM}$, the fraction of \lya flux transmitted through the IGM, under the assumption $T_\mathrm{IGM}$ does not depend on \lya or UV luminosity.

However, due to the inhomogeneous nature of reionization \citep[e.g.,][]{MiraldaEscude2000,Ciardi2003,Furlanetto2005,Mesinger2016}, the transmission fraction is in reality a broad distribution, which is not captured by the luminosity density estimates. Importantly, \citet{Mason2018} demonstrated the transmission fraction depends on not only the the average neutral fraction of hydrogen in the IGM, but also the galaxy's local environment and emission properties. For example, UV-bright galaxies have a higher transmission fractions at all neutral fraction values because their \lya line profiles are typically redshifted far into the damping wing absorption profile and they also typically exist in large reionized bubbles early in reionization \citep{Mason2018,Whitler2020}. By contrast, UV-faint galaxies emit \lya closer to their systemic velocity, which is thus more significantly absorbed by surrounding neutral IGM. They can be found in under-dense regions of the cosmic web where the IGM is still neutral, resulting in a lower transmission fraction even for high average neutral fractions.

This work models the evolution of the \lya LF as a function of the volume average neutral hydrogen fraction, \xHI, and redshift, z, to interpret observations and constrain reionization. We create our model by convolving the UV LF  with the \lya luminosity probability distribution as a function of \MUV. The models in this project include realistic, inhomogeneous simulations for reionization, enabling us to include the full distribution of \lya transmissions. This is an improvement on previous work which interpreted the \lya LF using fixed \lya\ transmission fractions \citep[e.g.,][]{Konno2018,Hu2019}, which may be considered an oversimplification. Furthermore, we use an analytic approach that enables us to model the \lya LF as a function of \xHI and $z$ independently, rather than using a simulation with a fixed reionization history \citep[e.g.,][]{Itoh2018} -- allowing us to disentangle the impact of IGM and redshift evolution.

This paper is structured as follows. In Section~\ref{sec:mod} we describe our model for the \lya LF. In Section~\ref{sec:results} we describe our results for the \lya LF and the evolution of the Schechter function parameters and luminosity density. We infer the evolution of the neutral fraction for $z>6$ by fitting our model to observations and we forecast predictions for future surveys with the Nancy Grace Roman Space Telescope, Euclid, and the James Webb Space Telescope. In Section~\ref{sec:disc} we discuss our results and we present our conclusions in Section~\ref{sec:conc}. 

We use the \cite{Planck2015} cosmology and all magnitudes are given in the AB system.

\section{Methods}\label{sec:mod}

Here, we describe the components of our model. In Section~\ref{sec:mod_LyaLF}, we describe the methodology used to model the \lya LF. Both model components – the \lya luminosity probability distribution and the UV LF – are described in the succeeding Sections~\ref{sec:mod_lumpd} and~\ref{sec:mod_UVLF}. Section~\ref{sec:mod_normLF} describes the normalization of the \lya LF. Section~\ref{sec:mod_bayes} describes the Bayesian framework used to infer the neutral fraction given the \lya LF model and observational data. In Section~\ref{disc_diffLya} we discuss the differences in observational datasets that led to omitting or including certain surveys in our analysis.

\subsection{Modelling the \lya luminosity function} \label{sec:mod_LyaLF}
We model the evolution of the \lya LF as a function of redshift and \xHI by convolving models for the \lya emission from Lyman-break galaxies (LBGs) and the UV LF during reionization. This enables us to disentangle the effects of redshift evolution from the evolution due to IGM absorption \citep[][]{Mason2015,Mesinger2016,Mason2018}.

The luminosity function (LF) of galaxies shows the number density of galaxies in a certain luminosity interval and is typically described using the \citet{Schechter1976} function: 
\begin{equation} \label{eqn:schechter_LF}
    \phi(L)\dd L = \phi^* \left(\frac{L}{L^*}\right)^\alpha \exp{\left(-\frac{L}{L^*}\right)} \, \dd \left(\frac{L}{L^*}\right)
\end{equation}
where $\phi^*$ is the normalization constant, $\alpha$ is the power law for faint-end slope for $L < L^*$, and there is an exponential cutoff at $L > L^*$. These parameters are known to be conditional on the observed wavelength and cosmic time, as well as the type of galaxy \citep[e.g.,][]{Dahlen2005}. The rest-frame UV LF has been measured in detail out to $z\sim10$ and is one of our best tools for studying the evolution of galaxy populations \citep[e.g.,][]{Bouwens2015a,Bouwens2015b, Finkelstein2015a, Oesch2015}.

Following \citet{Dijkstra2012,Gronke2015} we can predict the number density of LAEs by using the UV LF and \lya luminosity probability distribution for LBGs to model the \lya LF:
\begin{equation} \label{eqn:LAE_LF}
\begin{split}
 \phi_\mathrm{LAE}&(L_\alpha,\xHI,z)\dd L_\alpha = \dd L_\alpha F  \\ 
& \times\int_{\mathrm{M}_\mathrm{UV,min}}^{\mathrm{M}_\mathrm{UV,max}} \dd \MUV \, \phi(\MUV,z) P(L_\alpha \,|\, \MUV,\xHI,z) 
\end{split}
\end{equation}

Here, $\phi(\MUV,z) \dd \MUV$ is the UV LF in the range $\MUV \pm \mathrm{d}\MUV/2$ and is described in Section~\ref{sec:mod_UVLF}. $P(L_\alpha \,|\, \MUV,\xHI)$, is the conditional probability of galaxies that have a \lya luminosity $L_{\alpha}$ in $L_{\alpha} \pm \mathrm{d}L_{\alpha}/{2}$, given a \MUV value and neutral fraction and is described in Section~\ref{sec:mod_lumpd}. We integrate our \lya LF over the \MUV range $-24 < \MUV < -12$ covering the observed range of the UV LF, $-23 < \MUV < -17$.

The factor $F$ in Equation~\ref{eqn:LAE_LF} is a normalization constant to fit the LF model to observations and can be thought of as the ratio of predicted LAEs versus the total number of LAEs recorded. If the \lya luminosity distribution, $P(L_\alpha \,|\, \MUV,\xHI,z) $ accurately describes the luminosities of the same Lyman-break galaxies measured in the UV LF this factor should be $F = 1$ (See Section~\ref{sec:disc_F} for further discussion).

\subsubsection{\lya luminosity probability distribution}\label{sec:mod_lumpd}

The probability distribution for \lya luminosity is derived from the \lya rest-frame equivalent width (EW) probability density function $P(EW \,|\, \MUV)$ where $P(L_\alpha \,|\, \MUV,\xHI) = P(EW \,|\, \MUV,\xHI) \frac{\partial EW}{\partial L_\alpha}$. We use the rest-frame EW probability distribution models by \citet{Mason2018} (based on observations by \citealp{DeBarros2017}) who forward-model observed EW after transmission through 1.6 Gpc$^3$ inhomogeneous reionization simulations \citep{Mesinger2016} at fixed average neutral fraction $\xHI = 0.01 - 0.95$, with a spacing of $\Delta\xHI \sim 0.01-0.03$. We use the following relationship between \lya luminosity in $\mathrm{erg\,s}^{-1}$ and EW to obtain $\partial EW/\partial L_\alpha$:
\begin{equation}
    L_\alpha = EW \times L_{\textsc{uv},\nu} \times \frac{c}{\lambda_\lya^2}\left(\frac{\lambda_\lya}{\lambda_\textsc{uv}}\right)^{\beta +2}
\end{equation}
Here, $\lambda_\alpha \sim 1216$\,\AA\ is the wavelength of the \lya resonance, the rest-frame wavelength of the UV continuum is typically measured at $\lambda_\textsc{uv} \sim 1500$\,\AA\ , $\beta$ is the UV continuum slope, where we assume $\beta = -2$ typical for high redshifts \citep[e.g.,][]{Bouwens2014}, and $c$ is the speed of light. $L_{\textsc{uv},\nu}$ is the UV luminosity density:
\begin{equation}
L_{\textsc{uv},\nu} = 4\pi (10)^2 \times 10^{-0.4(\MUV + 48.6)} \, \mathrm{erg\,s}^{-1}\,\mathrm{Hz}^{-1}.
\end{equation}
We normalize $P(L_\alpha \,|\, \MUV,\xHI)$ over the luminosity range $L_\alpha = 0 - 10^{44.5}$ erg s$^{-1}$ to encompass a large \lya luminosity interval, and within our defined \MUV range between $-24 < \MUV < -12$ (see Section~\ref{sec:mod_LyaLF}).

\begin{figure}
\centerline{\includegraphics[width=0.49\textwidth]{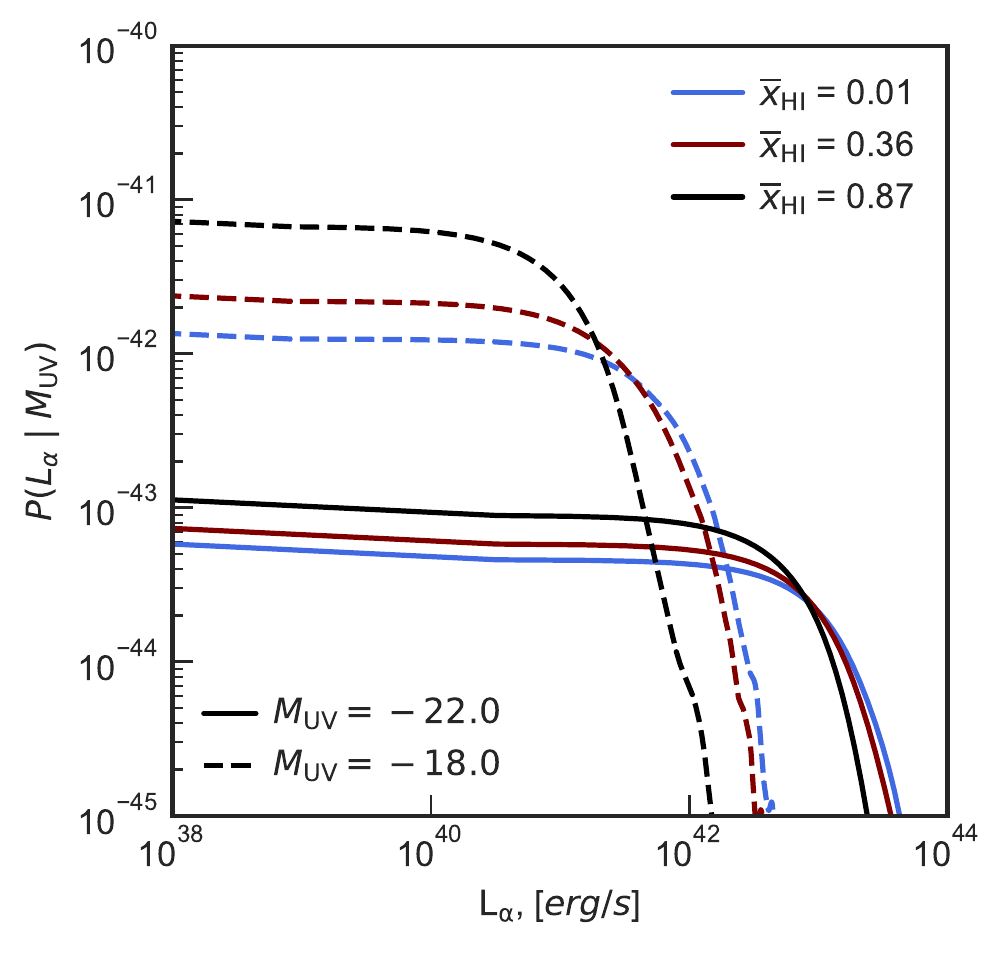}}
\caption{Example of the probability distribution for \lya luminosity, $P(\lya  \,|\,  \MUV, \xHI)$. This distribution is for UV-bright (solid line) and UV-faint galaxies (dashed line) at different \xHI values, $\xHI = 0.01, 0.36, 0.87$. When \xHI increases to an almost neutral environment, the probability of detecting luminous \lya galaxies decreases compared to in an almost fully ionized environment.}
\label{fig:LvsPL}
\end{figure}

The \lya luminosity probability distribution, $P(L_\alpha \,|\, \MUV, \xHI)$, is shown in Figure~\ref{fig:LvsPL} for a few \xHI and \MUV values. We note that as the \citet{Mason2018} models assume no evolution of the intrinsic \lya EW distribution with redshift -- the only redshift evolution is due to the increasing neutral fraction, our \lya luminosity distribution models also depend only on $\xHI$ with no additional redshift evolution. Recent works by \citet{Hashimoto2017,Jung2018,Shibuya2018} confirm a suitable approach to the \lya EW distribution models we incorporate into our work. Each paper ultimately notes no significant evolution in the \lya EW distribution with respect to redshift for $z\sim 5-7$.

Figure~\ref{fig:LvsPL} demonstrates large differences in the probability distribution for UV-bright and UV-faint galaxies. For UV-faint galaxies, we expect a higher probability of \lya luminosity emission at all \xHI values but with a \lya luminosity cut off at $L_\alpha \simlt  10^{42}$ erg s$^{-1}$. For UV-bright galaxies, at all \xHI we expect galaxies to have higher \lya luminosity values up to $L_\alpha \simlt 10^{44}$ erg s$^{-1}$, but this is rarer. For both UV-bright and UV-faint galaxies, as the neutral fraction increases, the probability of galaxies emitting strong \lya overall decreases.

The \MUV dependence of $P(L_\alpha \,|\, \MUV)$ is a direct consequence of the EW distribution model by \citet{Mason2018}. This model can be described as an exponential distribution plus a delta function:
\begin{equation} \label{eqn:EWdist}
\begin{split}
    P(EW | \MUV) =& \frac{A(\MUV)}{EW_c(\MUV)}e^{-\frac{EW}{EW_c(\MUV)}}H(EW) \\
    &+[1-A(\MUV)]\delta(EW)
    \end{split}
\end{equation}
where
\begin{align}
    A(\MUV) &= 0.65 + 0.1\,\mathrm{tanh}[3(\MUV+20.75)] \\
    EW_c(\MUV) &= 31 + 12\,\mathrm{tanh}[4(\MUV + 20.25)]\, \mathrm{\AA}
\end{align}

account for the fraction of emitters and the anti-correlation of EW with \MUV respectively. $H(EW)$ is the Heaviside step function and $\delta(EW)$ is a Dirac delta function (see Section 2.1.3 of \citealp{Mason2018} for further details). The intrinsic, emitted, distribution (i.e. $\xHI = 0$) is an empirical model fit to observations by \citet{DeBarros2017} at $z\sim 6$, where it was found that UV-bright galaxies had a lower probability of being emitters, and had lower average EWs \citep[consistent with previous findings; e.g.,][]{2006ApJ...645L...9A,Stark2010}. The model EW distribution is then painted onto galaxies in inhomogeneous reionization simulations, with different average neutral fractions, and the `observed' EW distribution in each of those simulations is recovered by sampling the transmission along thousands of sightlines. 

Although bright galaxies have low EW compared to faint galaxies, they are, on average, less affected by neutral gas in the IGM: UV-bright galaxies are typically more massive and reside in dense regions of the universe, in large IGM bubbles that have already reionized. In the simulations we use, reionization occurs first in overdense regions due to the excursion set formalism \citep[][]{Mesinger2007,Mesinger2016}. Observationally, clustering analyses show that UV-bright galaxies typically live in massive halos in dense regions \citep[e.g., Figure 15 of][]{Harikane2018}. So, \lya photons can escape more easily and EWs decrease at a slower rate (transmissions are already high). UV-faint galaxies have a higher intrinsic EW, on average, that decreases more rapidly than for UV-bright galaxies as the universe becomes more neutral. This is because they can be more typically found in neutral patches of IGM.

As described above in Section~\ref{sec:mod_LyaLF}, we generate our \lya LF by integrating the UV LF over the \MUV range $-24 < \MUV < -12$, covering the observed range of the UV LF, $-23 \lesssim \MUV \lesssim -17$. As the \citet{Mason2018} EW models were defined for $-23 \leq \MUV \leq -17$ to include galaxies outside of this range we set galaxies brighter than $\MUV = -23$ to have the same $P(L_\alpha  \,|\,  \MUV = -23)$, and all galaxies fainter than $\MUV =-17$ have the same $P(L_\alpha  \,|\,  \MUV= -17)$ (for a given $\xHI$).

\subsection{Galaxy UV luminosity functions} \label{sec:mod_UVLF}
In this paper, we use the \citet{Mason2015} UV LF model. In this model galaxy evolution is dependent on star formation that is associated with the construction of dark matter halos, with the assumption that these halos have a star formation efficiency that is mass dependent but redshift independent, which successfully reproduces observations over 13 Gyr \citep[see also, e.g.,][]{Trenti2010,Tacchella2013,Tacchella2018,Mirocha2020}.

\begin{figure}[t]
\centerline{\includegraphics[width=0.5\textwidth]{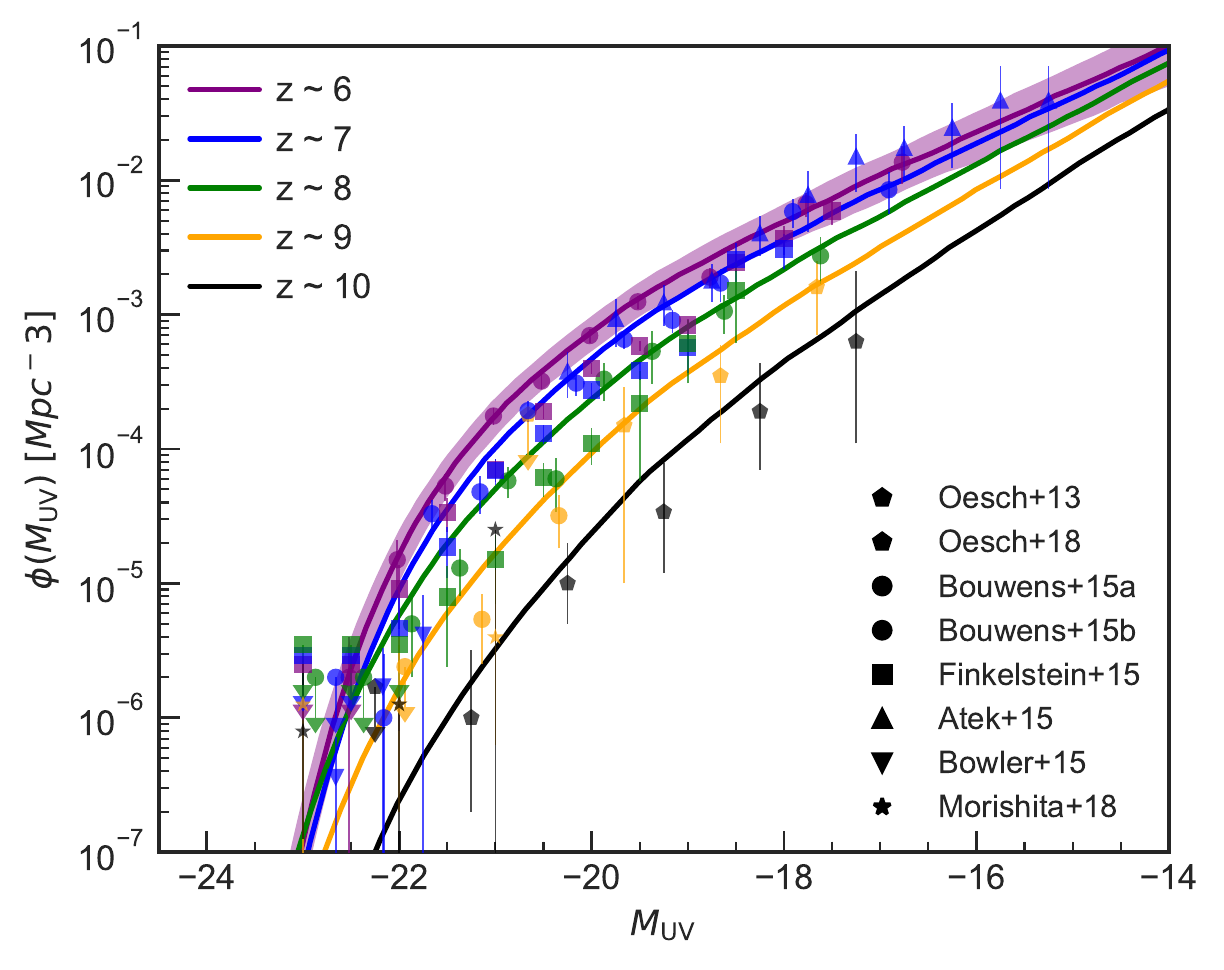}}
\caption{Modelled UV luminosity function for redshifts at $z =  6 - 10$ for UV-bright and faint galaxies, based on \citet{Mason2015} UV luminosity function model. There is a steep, exponential, drop-off for UV-bright galaxies that turns over into a power-law slope for UV-faint galaxies. The shaded region shows the $1\sigma$ confidence range at $z\sim6$ and similar regions can be assumed for each redshift. UV LF observations are shown with each marker color matching its corresponding UV LF model at a given redshift. Points show the binned UV LFs and upper limits from \citet{Oesch2013b,Oesch2018,Bouwens2015b,Bouwens2015a, Finkelstein2015b,Atek2015,Bowler2015,Morishita2018}.}
\label{fig:UVLF}
\end{figure}

The UV LF is plotted in Figure~\ref{fig:UVLF}. It is well-described by a \citet{Schechter1976} function (Equation~\ref{eqn:schechter_LF}).
The steep drop off of UV-bright galaxies can be explained in terms of rare high mass halos and their star formation efficiency: high mass halos are not efficient at forming stars, likely due to strong negative active galactic nuclei (AGN) feedback. The drop in number density with increasing redshift indicates a shift in star formation towards fainter, less massive galaxies.

\subsection{Calibrating the normalization of the \lya LF}\label{sec:mod_normLF}

To obtain an accurate model of the \lya LF to compare with observations, we must calibrate the \lya LF by finding the normalization constant, $F$. This factor accounts for any over-prediction in the number density of LAEs caused by the \lya luminosity distribution, $P(L_\alpha \,|\, \MUV,\xHI,z)$ \citep{Dijkstra2012,Gronke2015}. If the \lya luminosity distribution accurately describes the luminosities of the Lyman-break galaxies measured in the UV LF we should obtain $F = 1$.

We estimate $F$ using a maximum-likelihood approach to fit our model at $z = 5.7, \xHI \sim 0$ to the \citet{Konno2018, Ouchi2008} observations at $z = 5.7$. We set calibration at this redshift and neutral fraction as it is likely to be after the end of reionization \citep[e.g.,][]{McGreer2015}. Note, due to our reionization simulation grid (see Section~\ref{sec:mod_lumpd}), we use $\xHI=0.01$ for calibration, rather than $\xHI=0.0$, but note that the difference in $P(L_\alpha | \MUV, \xHI)$ should be negligible for such a small change in neutral fraction \citep[as shown by][]{Mason2018}.

We maximise the likelihood for the observed \lya LFs in each luminosity bin $L_i$: $P(\phi_i  \,|\,  \theta = F, L_i, \sigma_i)$ given our model $\phi_\mathrm{mod}(\theta=F, L)$. This estimation using binned LFs may not be the most optimal: a more accurate likelihood would be obtained using individual source information and the survey selection function \citep[see e.g.,][]{Trenti2008,Kelly2008,Schmidt2014,Mason2015a}, however collating of these data is not feasible within the scope of this project, and we leave this for future works. We note that \citet{Trenti2008} demonstrated that LFs estimated from binned data are generally in good agreement with those measured from unbinned data, but can bias the faint end slope towards steeper values. However, as the observed high-redshift \lya LFs are mostly $\simgt L^*$, we do not expect this to have a large impact on our results as the faint-end will already have large uncertainties.

Following \citet{Gillet2020} we use a split-norm likelihood (Equation~\ref{eqn:splitnorm}), to take into account asymmetric error bars. Assuming each observation and luminosity bin are independent, the total likelihood is: 
\begin{equation} \label{eqn:LF_likelihood}
P(\phi_{obs}  \,|\,  \theta = F, L) = \prod_i S(\phi_\mathrm{mod}(F, L_i), \mu_i, \sigma_{1,i}, \sigma_{2,i})
\end{equation}

Here, $\phi_\mathrm{mod}(F,L_i)$ is the model LF at luminosity $L_i$ for parameter $\theta = F$, $\mu_i$ is the observed number density value at $L_i$, and $\sigma_{1,i}$ and $\sigma_{2,i}$ are the respective lower and upper errors of the observed number density. In a single luminosity bin:
\begin{align} \label{eqn:splitnorm}
  S(\phi_\mathrm{mod}) &=
  \begin{cases}                                   B\exp[-\frac{1}{2}\frac{(\phi_\mathrm{mod}-\mu)^2}{\sigma_1^2}] & \text{if $\phi_\mathrm{mod} \leq \mu $} \\ B\exp[-\frac{1}{2}\frac{(\phi_\mathrm{mod}-\mu)^2}{\sigma_2^2}] & \text{if $\phi_\mathrm{mod} \geq \mu $} \\
  \end{cases}\\
  B &= \left[\sqrt{2\pi}\left(\frac{\sigma_1 + \sigma_2}{2}\right)\right]^{-1}
\end{align}

We minimize the logarithm of the likelihood Equation~\ref{eqn:LF_likelihood} to find $F$ using the Python package \verb|SciPy minimize|. The obtained minimum value is $F=0.974$ which we then use for \lya LF at all redshifts and \xHI values. 

Our recovered value of $F \approx 1$ indicates our luminosity distribution is a good model for Lyman-break galaxies. Other works such as \citet{Dijkstra2012,Gronke2015} found $F \sim 0.5$ at $z = 5.7$, and $F < 1$ at all lower redshifts, which is due to the different EW distribution they employed. We discuss this further in Section~\ref{sec:disc_F}.

\subsection{Bayesian Inference of the neutral fraction}
\label{sec:mod_bayes}
In Section~\ref{sec:results_xHI} we use our model to infer the IGM neutral fraction from observations. 

Bayes' theorem allows us to establish a posterior distribution for \xHI given observations. Bayes' theorem is defined as:
\begin{equation} \label{eqn:bayes}
\begin{split}
    P(\xHI \,|\,\{ \phi_{\mathrm{obs},i}(L_i)\}, z) = \\ \frac{P(\{ \phi_{\mathrm{obs},i}(L_i)\} \,|\,  \xHI, z) P(\xHI | z)}{P(\{ \phi_{\mathrm{obs},i}(L_i)\})}
    \end{split}
\end{equation}
Here, $\{\phi_{\mathrm{obs},i}(L_i)\}$ is the set of observed data in luminosity bins $L_i$ (where $\mu_i$ is the observed number density value at $L_i$, and $\sigma_{1,i}$ and $\sigma_{2,i}$ are the respective lower and upper errors of the observed number density). $P(\{ \phi_{\mathrm{obs},i}(L_i)\} \,|\,  \xHI, z)$ is the likelihood of obtaining our observed data given the model. $ P(\xHI | z) = P(\xHI)$ is the prior for the model parameter, $\xHI$, where we assume the neutral fraction is independent of redshift. More physically, this prior could be dependent on redshift, but we leave the prior independent of redshift to allow more flexibility when estimating the neutral fraction. Regardless, we still see the inferred neutral fraction increase with redshift. We use a uniform prior from 0 to 1. Although this is technically not needed, making our approach essentially a maximum likelihood estimate, we keep the Bayesian formalism to allow more physical priors to be used in future works. $P(\{ \phi_{\mathrm{obs},i}(L_i)\})$ is the Bayesian information that normalizes the posterior distribution.

We obtain the posterior distribution of the neutral fraction of hydrogen using our \lya LF model given the observed \lya luminosity values, $L_{\alpha}$, number density, $\phi(L_{\alpha})$, and the uncertainties in the number density. We use the same split-norm likelihood defined in Equation~\ref{eqn:LF_likelihood} with $\phi_\mathrm{mod}(\xHI, L_i)$. Further explanation of the inference of the neutral fraction can be seen in Appendix~\ref{appendix_xHI}.

We include uncertainties in our model \lya LF due to uncertainties in the UV LF via a Monte Carlo approach. We generate 100 UV LFs with a $0.2$ dex uncertainty in number density \citep[estimated from the][UV LF model]{Mason2015}. We then calculate the standard deviation of the resulting \lya LF, $\sigma_\mathrm{mod}$, as a function of \lya luminosity. We find the standard deviation is well-described by $\sigma_\mathrm{mod}(L_\alpha) \approx 0.1 \times \phi_\mathrm{mod}(L_\alpha)$. We use this uncertainty in calculating the likelihood (Equation~\ref{eqn:splitnorm}) where: 
\begin{align}
    \sigma_1 &\rightarrow \sqrt{\sigma_1^2 + \sigma_\mathrm{mod}^2} \\
    \sigma_2 &\rightarrow \sqrt{\sigma_2^2 + \sigma_\mathrm{mod}^2}
\end{align}
%

\subsection{\lya LF observational datasets}
\label{disc_diffLya}

In comparing our model to observations, we wanted to ensure we used datasets where the selection strategies were similar to each other and similar to the datasets used to calibrate our model (Section~\ref{sec:mod_normLF}), as it is known that different survey selection techniques can produce different estimates of the \lya LF \citep[for more discussion see][]{Taylor2020}. This led to the inclusion or exclusion of certain surveys from the estimation of the neutral fraction. In general, we aimed to use surveys which covered the widest areas (to minimize cosmic variance) and deepest \lya luminosity limits.

For the neutral fraction inference (Section~\ref{sec:results_xHI}) it was important to use observed LFs that were calculated consistently with each other and our calibration LF at $z=5.7$ (Section~\ref{sec:mod_normLF}). As \citet{Konno2018} covers the largest area, we used their LF for our calibration. Therefore, for the neutral fraction inference, we included additional datasets which covered the largest redshift range with similar flux measurements and corrections for their systematic uncertainties.

The observational datasets we used to infer the neutral fraction, and their survey areas, are as follows: \lya LFs at $z = 5.7,6.6$ by \citet{Konno2018} who surveyed $\sim 13.8$ and $ \sim 21.2 \, \mathrm{deg}^2$ areas of the sky using Subaru/Hyper Suprime-Cam (HSC) Subaru Strategic Program (SSP) Survey for redshifts $z = 5.7, 6.6$ respectively, and by \citet{Ouchi2008,Ouchi2010} who surveyed a $1 \, \mathrm{deg}^2$ area of the sky using Subaru/\textit{XMM-Newton} Deep Survey (SXDS) fields for both redshifts $z = 5.7, 6.6$. At $z = 7.0$, we used \lya LFs observed by \citet{Ota2017} who measured the total effective area of the Subaru Deep Field (SDF) and SXDS survey images for LF candidates to be $\sim0.5\, \mathrm{deg}^2$. \citet{Itoh2018} conducted an ultra-deep and large-area HSC imaging survey under the Cosmic HydrOgen Reionization Unveiled with Subaru (CHORUS) Program in a total of $3.1 \, \mathrm{deg}^2$ using two independent blank fields. Finally, \citet{Hu2019} implemented a large area survey using the Lyman Alpha Galaxies in the Epoch of Reionization (LAGER) project's deep-fields COSMOS and  Chandra Deep Field South (CDFS) covering an effective area of $2.14 \, \mathrm{deg}^2$. \lya LFs $z = 7.3$ are identified by \citet{Konno2014}, who surveyed a $\sim 0.5 \, \mathrm{deg}^2$ area in the SXDS and COSMOS fields and \citet{Shibuya2012}, who surveyed a total area of $1719\, \mathrm{arcmin}^2$ ($\sim0.5 \, \mathrm{deg}^2$) in the SDF and the Subaru/XMM-Newton Deep Survey Field (SXDF), using the Suprime-Cam. 

We ultimately excluded LFs measured by \citet{Santos2016} at $z = 5.7, 6.6$ when estimating the neutral fraction because these LFs were significantly higher than those by \citet{Konno2018, Ouchi2008, Ouchi2010}. This is most likely due to differences in incompleteness corrections and the methodology for taking \lya flux measurements from narrow-band images \citep[see][for more discussion]{Santos2016,Konno2018}. We also excluded the LFs measured by \citet{Taylor2020} at $z = 6.6$ from the estimation of the neutral fraction, where unlike other works, they corrected for an selection incompleteness, however our model only incorporates observations un-corrected for selection incompleteness (this decision is discussed further in Section~\ref{sec:results_evo_LyaLF}).

Although there are \lya LF measurements at higher redshift values \citep[e.g.,][at $z = 7.7$]{Hibon2010,Tilvi2010,Krug2012,Clement2012, Matthee2014}, we decide not to include these works in comparison to our model. Ultimately, the areas of surveys greater than $z = 7.3$ are much smaller than surveys completed at lower redshifts. Therefore, surveys at $z > 7.3$ more likely to be biased because reionization is inhomogeneous \citep[see, for example, Figure 11 from][where they compare LFs for different survey areas]{Jensen2013}. The highest redshift LAE candidates are also prone to higher rates of contamination \citep[][]{Matthee2014}, so with only the inclusion of lower redshifts, we can obtain more robust estimations of the neutral fraction.

\section{Results}\label{sec:results}

In this Section, we describe the evolution of our model for the \lya LF.
In Section~\ref{sec:results_predLF}, we describe the expectation value of \lya luminosity at a given \MUV to understand what region of the \lya LF galaxies from a given UV magnitude range dominate. In Section~\ref{sec:results_evo_LyaLF}, we present our predicted \lya LF and compare with observations. In Section~\ref{sec:results_evo_schech}, we describe the evolution of the Schechter parameters for our \lya LF model from $z=5-10$. In Section~\ref{sec:results_LD}, we show our results for the \lya luminosity density as a function of redshift and \xHI. In Section~\ref{sec:results_xHI}, we present our inference of the IGM neutral fraction. Section~\ref{sec:results_pred}, presents predictions for future surveys with the Nancy Grace Roman Space Telescope, Euclid, and the James Webb Space Telescope from our model.

\subsection{The average \lya luminosity of LBGs}
\label{sec:results_predLF}

\begin{figure}[t]
    \centering
    \includegraphics[width=0.49\textwidth]{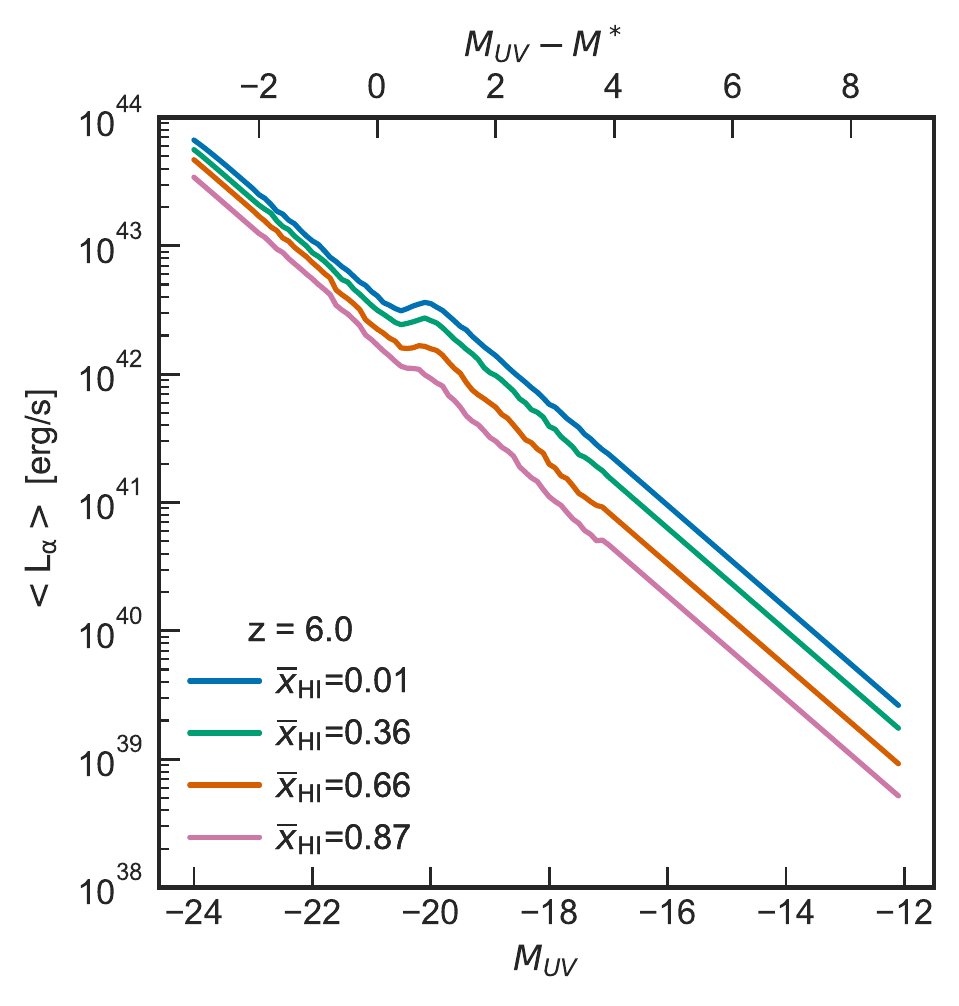}
    \caption{Expectation value for \lya luminosity at $z=6.0$ for a range of \MUV values and \xHI values. Here, we also show how $\langle L_\alpha \rangle$ compares for galaxies brighter or fainter than $\MUV^*$ (where we use $\MUV^* = -20.9$ at $z= 6.0$ from \citealp{Mason2015}). The model expects UV-faint galaxies to have an average \lya luminosity lower than that of UV-bright galaxies. As the neutral fraction increases, $\langle L_{\alpha} \rangle$ decreases. For UV-bright galaxies, there is not much decrease in the \lya luminosity expected (a factor of $\sim 2$), compared to a factor $\sim10$ for UV-faint galaxies. The bump in the plot, around $\MUV \sim -20$, is due to the EW probability distribution threshold between UV-bright and UV-faint galaxies \citep[see][]{Mason2018}, and is discussed further in Section~\ref{sec:results_predLF}.}
    \label{fig:expecLya}
\end{figure}

To understand the impact of environment and galaxy properties on the evolution of the \lya LF during reionzation, we investigate the typical \lya luminosity of LBGs. 
In Figure~\ref{fig:expecLya} we plot the expectation value of \lya luminosity, $\langle L_{\alpha} \rangle$, as a function of UV magnitude at $z= 6.0$. The expectation value is defined as:
\begin{equation} \label{eqn:L_expect}
    \langle L_{\alpha} \rangle = \int_{L_{\mathrm{min}}}^{L_{\mathrm{max}}}{L_\alpha \cdot P_{\mathrm{norm}}(L_\alpha \,|\, \MUV) \, \dd L_\alpha}
\end{equation}
\begin{equation} \label{eqn:pnorm}
    P_{\mathrm{norm}}(L_\alpha \,|\, \MUV) = \frac{P(L_\alpha \,|\, \MUV)}{\int_{L_{\mathrm{min}}}^{L_{\mathrm{max}}}{P(L_\alpha \,|\, \MUV)}}
\end{equation}
where we calculate the integrals over the range $10^{36} < L_\alpha < 10^{44.5}$ erg s$^{-1}$. This range (i.e., a range greater than zero) is chosen because we want to observe the typical $\langle L_\alpha \rangle$ - \MUV relation for \lya emitters. We also want to ensure coverage of the \lya luminosity values over the \MUV range $-24 \leq \MUV \leq -12$.

Figure~\ref{fig:expecLya} demonstrates that for UV-bright galaxies, $\MUV \lesssim -20$, we expect an average \lya luminosity $L_{\alpha} \sim 10^{42} - 10^{43.6}$ erg s$^{-1}$. For UV-faint galaxies we expect a lower typical \lya luminosity of $L_{\alpha} \sim 10^{39} - 10^{41}$ erg s$^{-1}$. Here, we also show how $\langle L_\alpha \rangle$ compares for galaxies brighter or fainter than $\MUV^*$ (where we use $\MUV^* = -20.9$ at $z= 6.0$ from \citealp{Mason2015}).

Figure~\ref{fig:expecLya} shows a decrease in \lya luminosity for a given \MUV as \xHI increases, as expected due to the reduced transmission in an increasingly neutral IGM \citep{Mason2018}. This effect is strongest for UV-faint galaxies, where the average \lya luminosity decreases by a factor of $\sim10$ as \xHI increases to 1. UV-bright galaxies do not show much decrease in \lya luminosity at different \xHI. The more sizeable impact of reionization on UV-faint galaxies is because they typically exist in the outskirts of dense IGM environments. Thus, a more neutral IGM shifts their \lya luminosity towards even lower values. 

The bump in the plot, around $\MUV \sim -20$, is due to the EW probability distribution threshold between UV-bright and UV-faint galaxies \citep{Mason2018}.
We tested the importance of this bump by fixing the $P(EW)$ distribution (in our case we tested at $P(EW | \MUV) = P(EW | \MUV = -17)$) which removes the bump. However, this drastically affected the \lya LF model where it does not fit observations well on the bright-end. This means that $P(EW)$ \textit{must} be shifted to lower EW values for galaxies brighter than $\MUV < -20$.

\subsection{Evolution of the \lya \ luminosity function}
\label{sec:results_evo_LyaLF}

\begin{figure*}
\centering
  \includegraphics[width=0.49\textwidth]{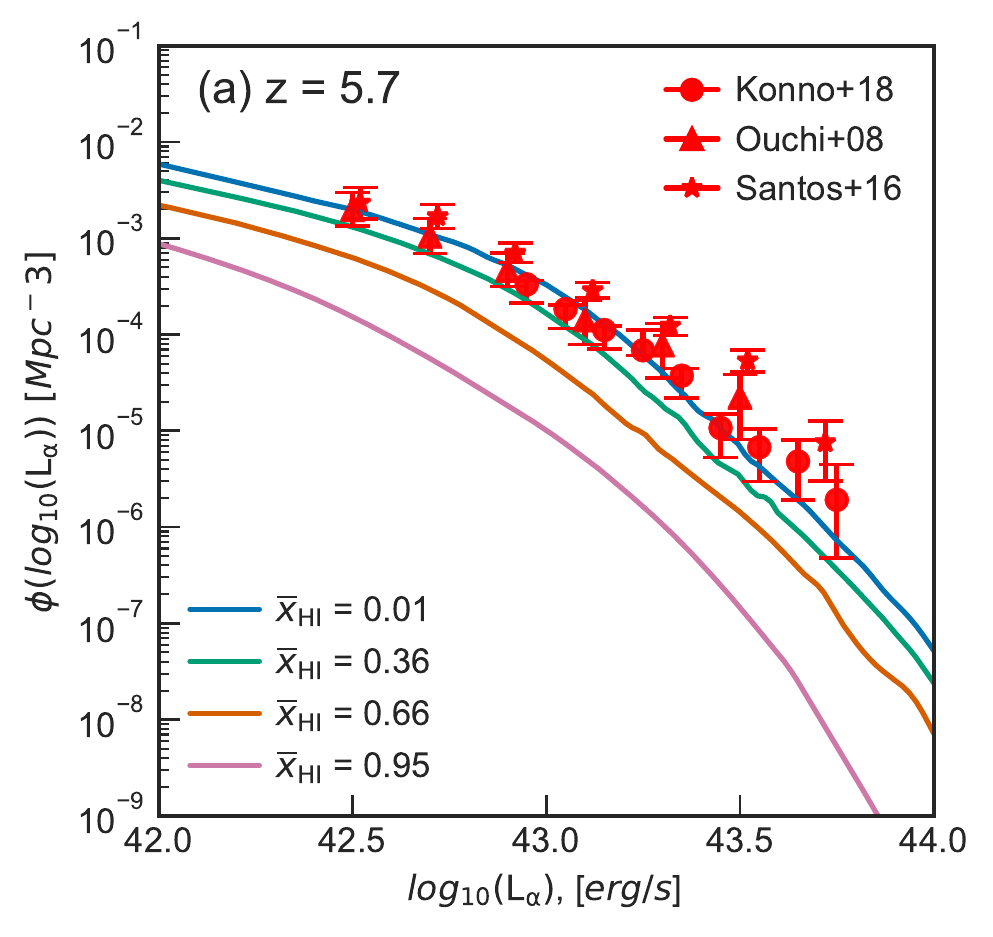}
  \includegraphics[width=0.49\textwidth]{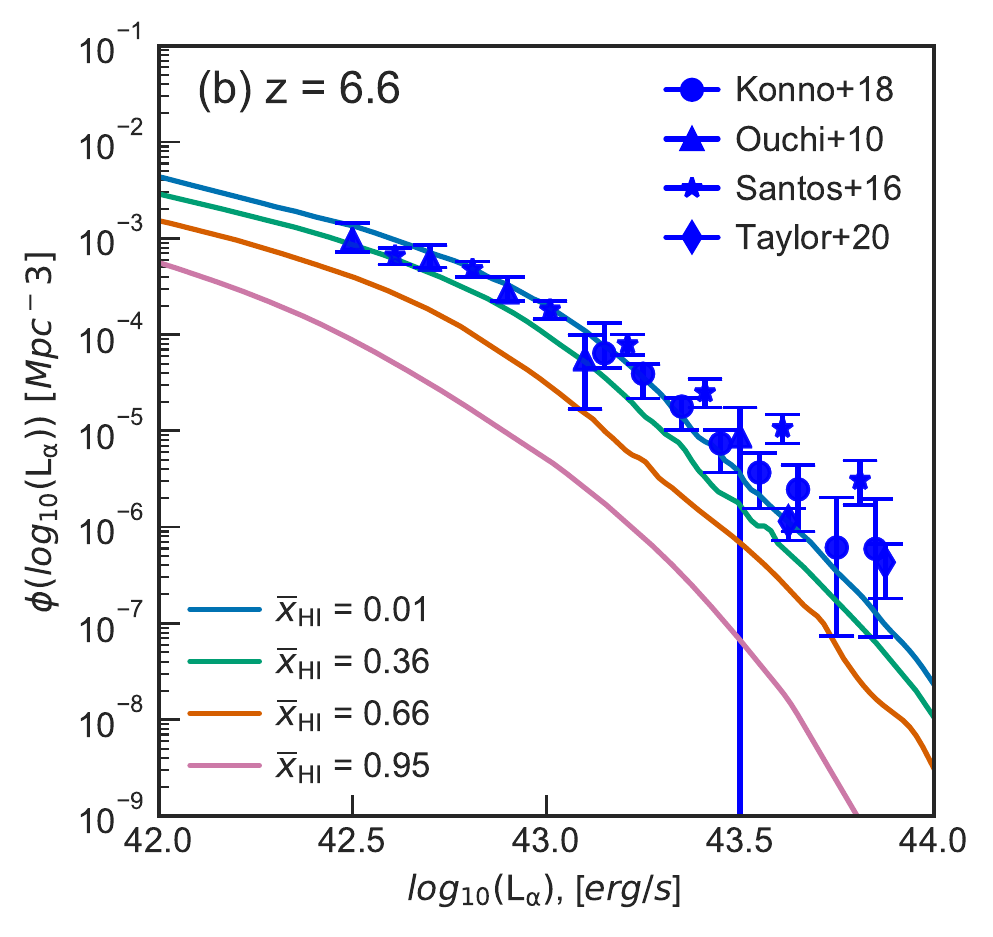}
  \includegraphics[width=0.49\textwidth]{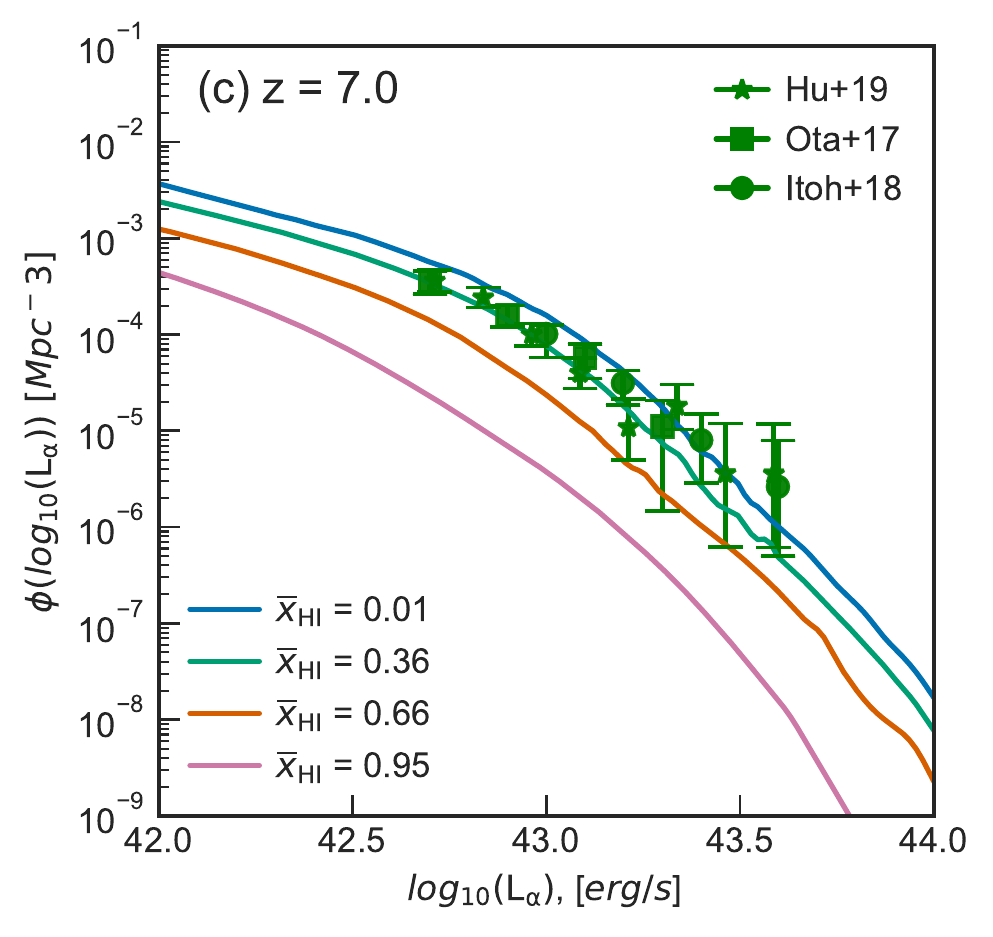}
  \includegraphics[width=0.49\textwidth]{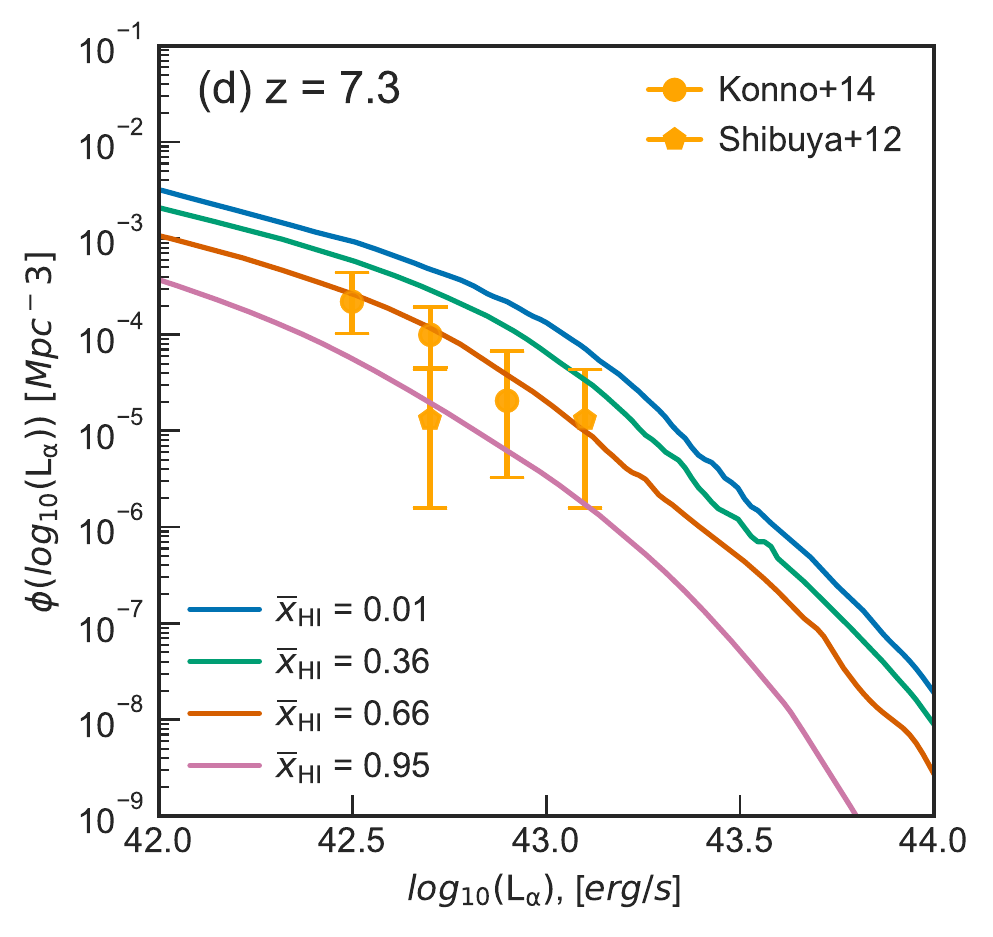} 
\caption{Our predictions for \lya LF for $z = 5.7, 6.6, 7.0, 7.3$ at $\xHI = 0.01 - 0.87$ (explained in Section~\ref{sec:results_evo_LyaLF}). We also plot observations by \citet{Ouchi2008,Ouchi2010,Shibuya2012, Konno2014,Santos2016, Ota2017,Konno2018,Itoh2018, Hu2019, Taylor2020} for comparison with our model. Each model line corresponds to a different neutral fraction. The \lya LF model decreases and changes shape at each redshift as \xHI increases. By comparing the observations to our model, as redshift increases, an increasingly neutral IGM is favoured.}
\label{fig:lyaLF}
\end{figure*}

We compare our model \lya LF to observations at $z = 5.7, 6.6, 7.0, 7.3$. In Figure~\ref{fig:lyaLF}, we plot our \lya LF models for a range of \xHI from a fully neutral to fully ionized IGM at a given redshift. We also plot observations by  \citet{Ouchi2008,Ouchi2010,Shibuya2012,Konno2014,Santos2016, Ota2017,Konno2018,Itoh2018, Hu2019, Taylor2020} for comparison. Note that we use the selection incompleteness un-corrected LFs by \citet{Hu2019} for the best comparison with other observations and our model, which is calibrated using data which do not account for this incompleteness. Our simple model reproduces the shape of the \lya LF remarkably well. Note that our \lya LF is slightly lower than the one observed by \citet{Santos2016}. This mismatch between their \lya LF and other  works found in literature is known and discussed, e.g., in \citet{Taylor2020, Hu2019}.

We see that at $z=5.7,6.6,7.0$ the observations are fairly consistent with $\xHI \sim 0.15 - 0.36$, whereas at $z = 7.3$ the data are more consistent with $\xHI \sim 0.66 - 0.87$, suggesting an increasingly neutral IGM environment as redshift increases, consistent with other observations at $z \sim 7$ \citep[e.g.,][]{Mason2018,Whitler2020,Mason2019,Hoag2019}.

Our model predicts that there is not much decrease in number density at low neutral fractions, $\xHI \simlt 0.4$ but the LF decreases more rapidly at higher neutral fractions. Based on Figure~\ref{fig:expecLya} the lack of evolution at $\xHI \simlt 40\%$ can be explained by the fact that the bright-end of the \lya LF is dominated by UV-bright galaxies which exist in over-dense regions of IGM that tend to reionize early \citep[][]{Mesinger2007,Mesinger2016,Harikane2018}. Thus, only in reionization's earliest stages do these galaxies experience significant reduction in transmission.

\subsection{Evolution of Schechter function parameters for the \lya LF}
\label{sec:results_evo_schech}

We fit Schechter parameters, $\alpha, L^{*}, \phi^{*}$, for our models using \verb|emcee| \citep{Emcee} to predict how the shape of the \lya LF evolves with redshift and neutral fraction. To compare with observations, we fit the LF over the luminosity range $42.5 < \log_{10}L_\alpha/\mathrm{erg\,s}^{-1} < 44$. We fit the Schechter function to our \lya LF models with all possible combinations of redshift and \xHI but point out that the resulting \lya LF are not exact Schechter functions for realistic \lya EW distributions (even if the input UV LF is) -- our \lya LFs are generally less steep at the bright-end than a Schechter function. Further details about the fitting are provided in Appendix~\ref{appendix_schechter}.

In Figure~\ref{fig:evo_schech} we plot the evolution of each parameter (where we show the median value from the fits) with respect to redshift and \xHI. We see that the parameters decrease overall as the universe becomes more neutral due to \lya photon attenuation and decrease with redshift because galaxies become fainter and rarer as redshift increases (as seen in the UV LF (Figure~\ref{fig:UVLF})).

The left panel of Figure~\ref{fig:evo_schech} shows redshift versus $\alpha$, which is the power law slope for very low luminosities. At fixed \xHI, $\alpha$ decreases as redshift increases, within the slope range of approximately $-2.5 < \alpha < -1.8$, as expected due the hierarchical build-up of galaxies producing an increasingly steep faint-end slope of the UV LF with redshift \citep{Mason2015}. The points plotted at each redshift show the impact of neutral hydrogen. $\alpha$ decreases significantly more due to the neutral gas than it does with redshift because \lya attenuation affects faint galaxies more and thus makes them fainter, forcing them further back into the \lya LF.

The center panel of Figure~\ref{fig:evo_schech} reveals that $L^*$ decreases in the range $z = 5 - 7$, but increases sharply toward $z = 8$, and declines toward higher redshifts at fixed neutral fractions. This upturn is a consequence of the evolving shape of the UV LF due to dust attenuation at these redshifts in our model. In \citet{Mason2015}, there is an overlapping between $z = 6 - 8$ for the UV LF model around $\MUV = -23 $ which corresponds to $L_\alpha = 10^{43.6} \, \mathrm{erg \,s}^{-1}$, also seen in observations \citep[e.g.,][]{Bouwens2015a,Bouwens2015b}. This overlapping is consistent with a reduction in dust obscuration, such that younger, brighter galaxies at higher redshifts contain less dust and so, there is a possibility of observing more of them and shifting the LF models towards higher luminosities. As the neutral fraction increases at each redshift, the characteristic \lya luminosity, $L^*$, decreases. This trend can be attributed to an increasing attenuation of \lya photons from UV-bright galaxies as the neutral fraction increases.

The right panel of Figure~\ref{fig:evo_schech} shows a decreasing number density of \lya emitting galaxies as we look back to higher redshifts. For each redshift, as shown, assuming the IGM is ionized, more \lya emitting galaxies are expected to be visible to us and thus the number density increases with decreasing redshift, compared to a neutral IGM. At higher redshifts and at a fixed neutral fraction, we see an overall decrease in number density of \lya emitters -- due to the overall reduction in the number of galaxies at high redshifts \citep[as seen in the evolution of the UV LF, e.g.,][]{Bouwens2015b, Bouwens2015a,Mason2015}.


We did not compare our Schechter function parameters directly with observations as previous works used a fixed $\alpha$ to determine their best-fit Schechter function parameters \citep[e.g,][]{Konno2014,Konno2018,Ouchi2008,Ouchi2010,Itoh2018,Ota2017,Hu2019}. As the Schechter function parameters are degenerate \citep[see, e.g., discussion in][]{Herenz2019}, it is difficult to compare with our model directly. Also note that our model for the \lya LF is not well described by a Schechter fit -- our model LFs are typically less steep at the bright end than a Schechter function's exponential drop-off (see Figure~\ref{fig:Emceeplots}). In our approach (Equation~\ref{eqn:LAE_LF}) we do not expect the \lya LF to be Schechter form as the integral of the Schechter UV LF does not have a Schechter form, as discussed in Section 3.1 of \citet{Gronke2015}. Physical reasons for an observed bright-end excess in the \lya LF are discussed in Section 4.1 of \citet{Konno2018}, including the contribution of AGN, large ionized bubbles around bright LAEs and gravitational lensing.

\begin{figure*}
\includegraphics[width=0.33\textwidth]{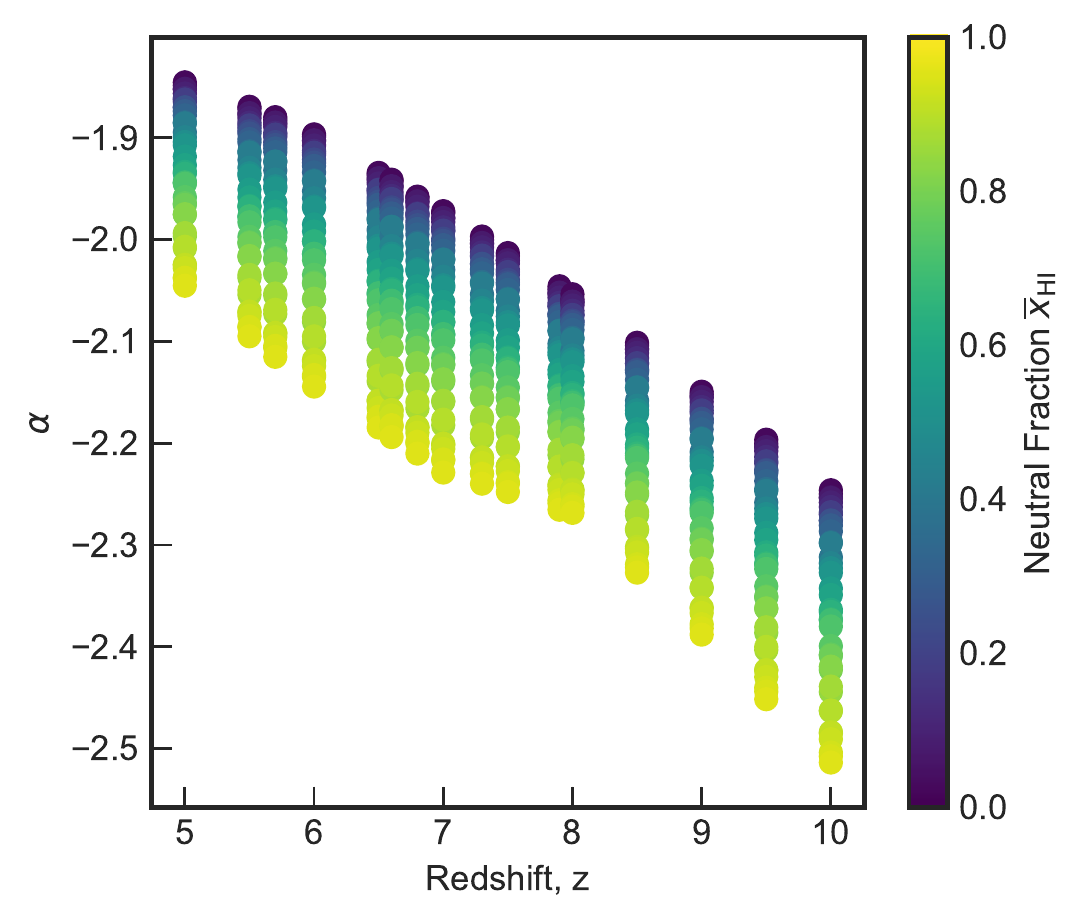}  
\includegraphics[width=0.33\textwidth]{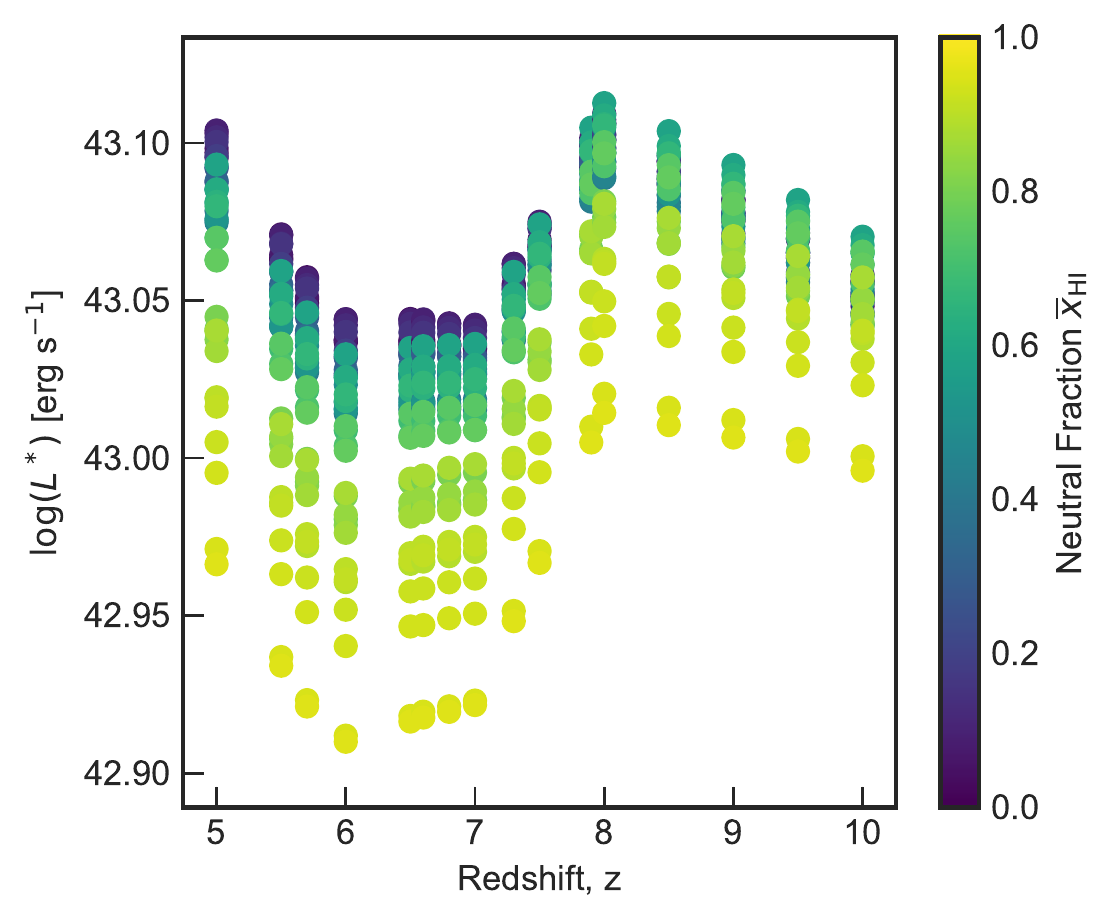}  
\includegraphics[width=0.33\textwidth]{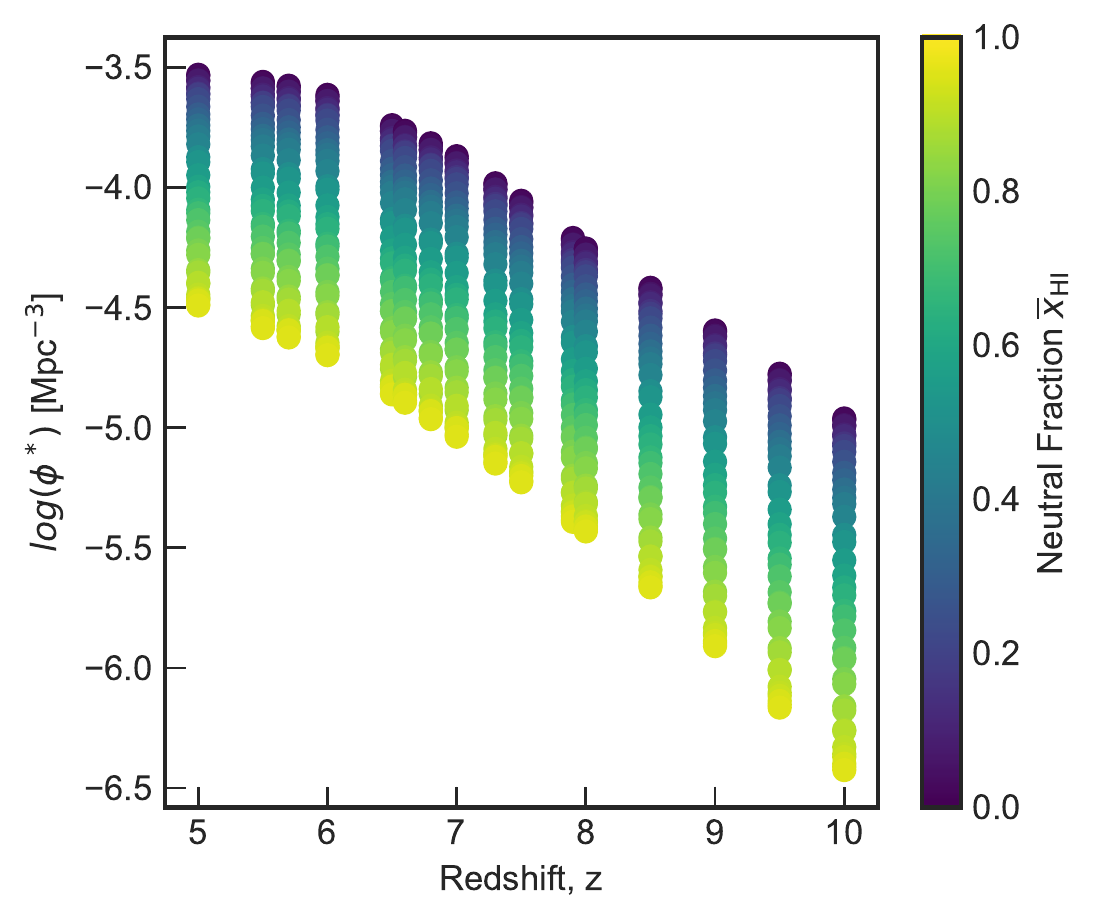}
\caption{The evolution of the Schechter function parameters $\alpha, L^{*}, \phi^{*}$ (left, center, right panels respectively), as a function of redshift $z = 5 - 10$ with a step of $\Delta z = 0.5$. Models at different average neutral fractions, \xHI are shown with different colors referenced on the adjacent colorbar. Further details about the fitting procedure are explained in Sections~\ref{sec:results_evo_schech} and Appendix~\ref{appendix_schechter}. We find strong evolution of the Schechter function parameters with \xHI, in which each parameter decreases as the neutral fraction increases because \lya photons become more attenuated.}
\label{fig:evo_schech}
\end{figure*}

\subsection{Evolution of \lya luminosity density}
\label{sec:results_LD}

The \lya luminosity density (LD) is the total energy emitted in \lya by all galaxies obtained by integrating the luminosity function:
\begin{equation} \label{eqLD}
    \rho_\alpha(z, \xHI) = \int L_\alpha \phi(L_\alpha,z,\xHI)\,\mathrm{d}L_\alpha
\end{equation}
where $\phi(L_\alpha,z,\xHI)$ is our \lya LF model number density. We generate $\rho_\alpha(z, \xHI)$ from our model by integrating over our luminosity grid $10^{42.4} \leq L_\alpha \leq 10^{44.5}$ erg s$^{-1}$ to compare with observations that use similar luminosity limits, as the luminosity density is highly sensitive to the minimum luminosity, due to the power-law slope of the LF faint end. 

\begin{figure}
    \centering
    \includegraphics[width=0.49\textwidth]{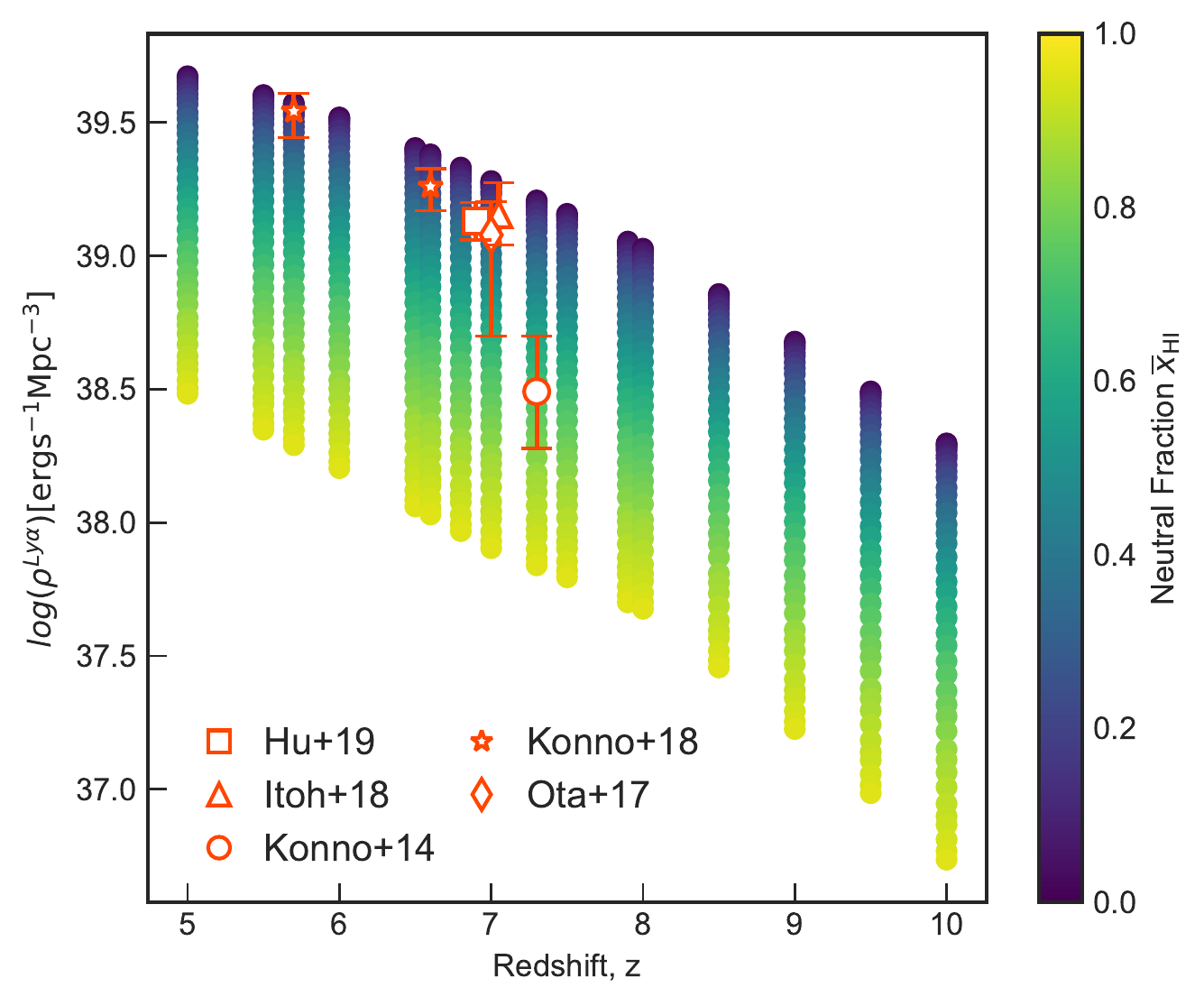}
    \caption{Evolution of the \lya luminosity density as a function of redshift and the IGM neutral fraction between $z = 5-10$. Observational data from \citet{Konno2018} (orange stars, $z = 5.7, 6.6$), \citet{Hu2019} (orange squares, $z = 6.9$), \citet{Itoh2018} (orange triangles, $z = 7.0$), \citet{Ota2017} (orange diamonds, $z = 7.0$) and \citet{Konno2014} (orange circles, $z = 7.3$), show a clear drop in the overall \lya luminosity density as redshift increases. The LD model shows a decreasing trend as redshift increases and we can expect to see lower LD values as the neutral fraction increases at any redshift.}
    \label{fig:LDplot}
\end{figure}

Figure~\ref{fig:LDplot} shows the evolution of the luminosity density along a range of redshifts, $z = 5 - 10$, and the neutral fraction, \xHI, predicted by our model. The modelled luminosity density decreases by a factor of $\sim 10$ as the universe becomes more neutral and declines overall as redshift increases. Observations from \citet{Konno2014,Konno2018,Hu2019,Itoh2018,Ota2017} at $z = 5.7, 6.6, 7.0, 7.3$ show a decrease in luminosity density to higher redshifts. We compare our model with observations and see that the LD observations are consistent with a mostly ionized IGM at $z = 5.7, 6.6, 7.0$ and increase towards a more neutral IGM at $z = 7.3$. These observations, similar to the \lya LF observations, are chosen based on their selection incompleteness being un-corrected further explained in \citet{Hu2019}. We also chose to plot \lya LD observations that were considered fiducial (some works also tested separate LD points at different $\alpha$ to compare with others).

\subsection{The evolution of the neutral fraction at $z>6$}
\label{sec:results_xHI}

We perform a Bayesian inference for the neutral fraction based on our model, as explained in Section~\ref{sec:mod_bayes}. We infer $\xHI$ at $z = 6.6, 7.0, 7.3$ by fitting our model to LF observations by \citet{Shibuya2012,Konno2014,Konno2018,Itoh2018,Ota2017,Hu2019}. We infer $\xHI(z=6.6) = 0.08^{+ 0.08}_{- 0.05}, \, \xHI(z=7.0) = 0.28 \pm 0.05$ and $\xHI(z=7.3)=0.83^{+ 0.06}_{- 0.07}$ (all errors are $1\sigma$ credible intervals). Appendix~\ref{appendix_xHI} shows the posterior distributions for \xHI at each redshift (Figure~\ref{fig:xHIminimization}). 

We also show the comparisons between the posterior distributions for the neutral fraction obtained using the \lya LD observations \citep[][see Section~\ref{sec:results_LD}]{Konno2014,Ota2017,Konno2018,Itoh2018,Hu2019}. As expected, we see larger uncertainties in the estimations of the neutral fraction from the \lya LD data due to including fewer data points compared to the LFs. We find $\xHI(z = 6.6) = 0.22^{+ 0.12}_{- 0.11} $, $\xHI(z = 7.0) = 0.25 \pm 0.08$, and $\xHI(z = 7.3) = 0.69^{+ 0.12}_{- 0.11}$. Using the full LF data thus not only enables us to infer neutral fractions that are more robust to non-uniform \lya transmission, but adds a statistical advantage over previous luminosity density methods by reducing the uncertainty on the neutral fraction (for further discussion, we refer the reader to Section~\ref{disc_compare}).

\begin{figure}[t]
    \centering
    \centerline{\includegraphics[width=0.49\textwidth]{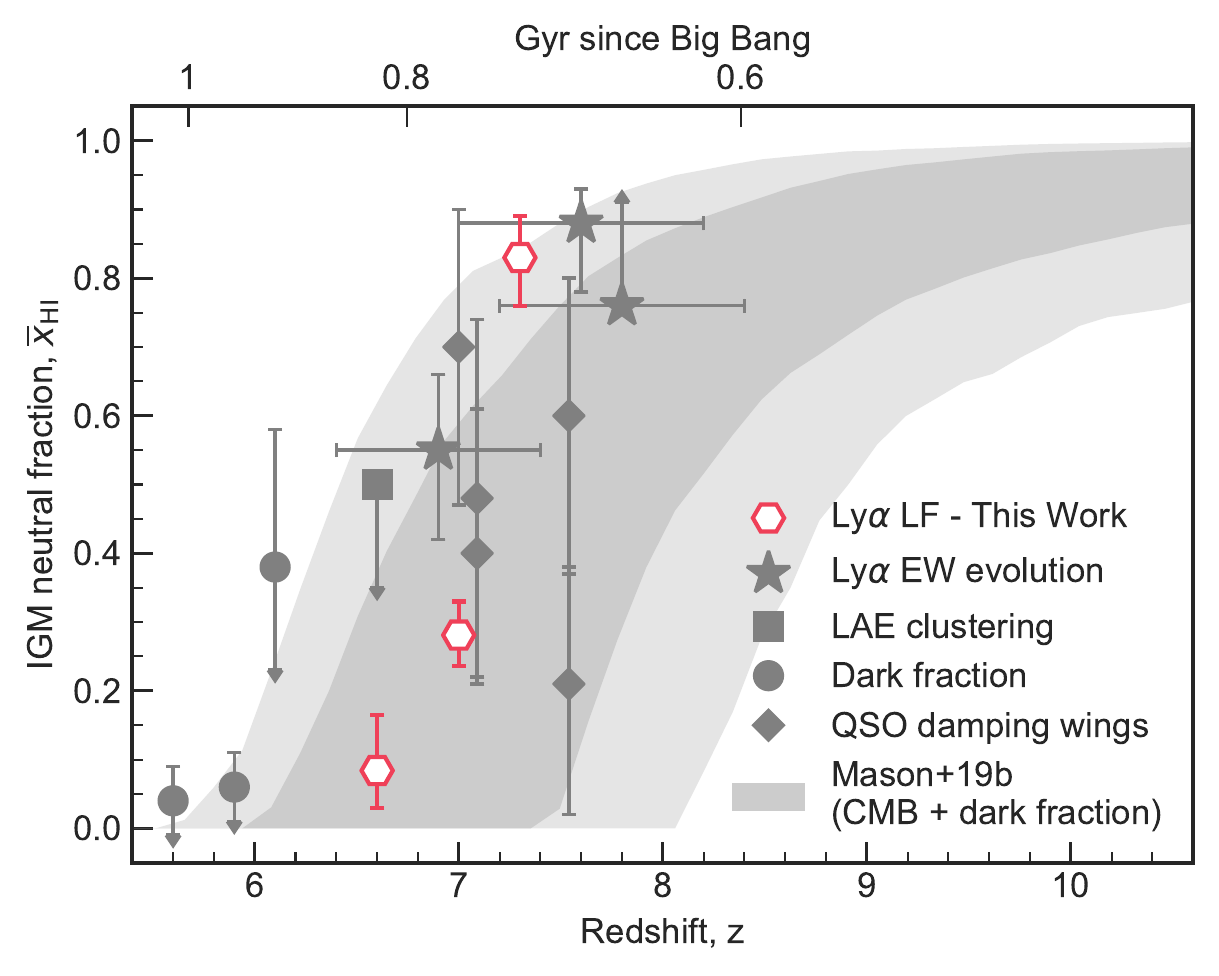}}
    \caption{IGM neutral fraction of hydrogen as a function of redshift updated from \citet{Mason2018}. Reionization history plot with this work corresponding to the \lya LF given a $1 \sigma$ uncertainty (red hexagons). Constraints derived from observations of previous estimates from the fraction of LBGs emitting \lya are plotted (star; \citet{Mason2018,Mason2019,Whitler2020, Hoag2019}); the clustering of \lya emitting galaxies (square; \citet{Ouchi2010}; \citet{Sobacchi2015}); \lya and Ly$\beta$ forest dark fraction (circle; \citet{McGreer2015}); and QSO damping wings (diamond; \citet{GreigMes2017}; \citet{Davies2018}; \citet{Greig2019}; \citet{Wang2020}).
    The shaded regions of the reionization plot show the corresponding $1\sigma$ and $2\sigma$ uncertainty coverage consistent with \citet{Planck2015} $\tau$ and dark fraction \citep{Mason2019b}.}
    \label{fig:reionization_history}
\end{figure}

Figure~\ref{fig:reionization_history} shows our new constraints on reionzation history along with other approaches to estimating the neutral fraction. Our results show clear upward trend in neutral fraction at higher redshifts, consistent with an IGM that reionizes fairly rapidly. Our measurement at $z=6.6$ is consistent with previous upper limits on the neutral fraction at $z\leq6.6$ \citep[][]{McGreer2015, Ouchi2010, Sobacchi2015}. Our measurement at $z=7.0$ is consistent with inferences from the \lya damping wing in the quasar ULAS J1120+0641 \citep{Davies2018}, but is lower than than constraints from the \lya EW distribution in LBGs by \citep{Mason2018,Whitler2020}, though is still consistent within $2\sigma$. Our measurement at $z=7.3$ is consistent with other constraints at $z>7$ \citep{Hoag2019,Mason2019,Davies2018} though, like the other $z>7$ constraints, is higher than the QSO damping wing measurements at $z = 7.5$ for the quasar ULAS J1342+0928 by \citet{Greig2019}.

\subsection{Predictions for Nancy Grace Roman Space Telescope, Euclid and JWST surveys}
\label{sec:results_pred}
The Nancy Grace Roman Space Telescope High Latitude Survey (NGRST HLS) and the Euclid Deep Field Survey (DFS) will both be particularly important surveys which will probe into higher redshifts and detect \lya emission lines further into the Epoch of Reionization with wide-area slit-less spectroscopy. The James Webb Space Telescope (JWST) will also be able to detect high redshift \lya with high sensitivity, albeit in smaller areas. Here we make predictions for potential \lya LFs with these telescopes.

The Euclid Deep Field Survey will cover a 40 sq. degree area at a $5\sigma$ flux limit of $\sim 8.6\times10^{-17}$ erg s$^{-1}$ cm$^{-2}$. It will cover $0.9 \leq \lambda_{\mathrm{obs}} \leq 1.3 \mu$m corresponding to a redshift range for \lya of $6 \leq z \leq 10$ \citep[e.g,][]{Laureijs2012,Bagley2017}. NGRST High Latitude Survey will cover $1.00 \leq \lambda_{\mathrm{obs}} \leq 1.95 \mu$m corresponding to a redshift range for \lya of $8 \leq z \leq 15$, and survey a 2200 sq. degree area at a $5\sigma$ flux limit of $\sim 7.1\times10^{-17}$ erg s$^{-1}$ cm$^{-2}$ \citep[e.g,][]{Ryan2019,Spergel2013}. JWST's slit-less spectrograph NIRISS will cover $0.7 \leq \lambda_{\mathrm{obs}} \leq 5.0 \mu$m, capable of detecting \lya at $z \gtrsim 5$. While there is no dedicated wide-area survey with NIRISS, we investigate a mock pure-parallel survey of 50 pointings ($\sim240$ sq. arcmin) with a $5\sigma$ flux limit of $\sim 5.0\times10^{-18}$ erg s$^{-1}$ cm$^{-2}$, assuming 2 hour exposures with the F115W filter.

\begin{figure}[t]
    \centering
    \centerline{\includegraphics[width=0.49\textwidth]{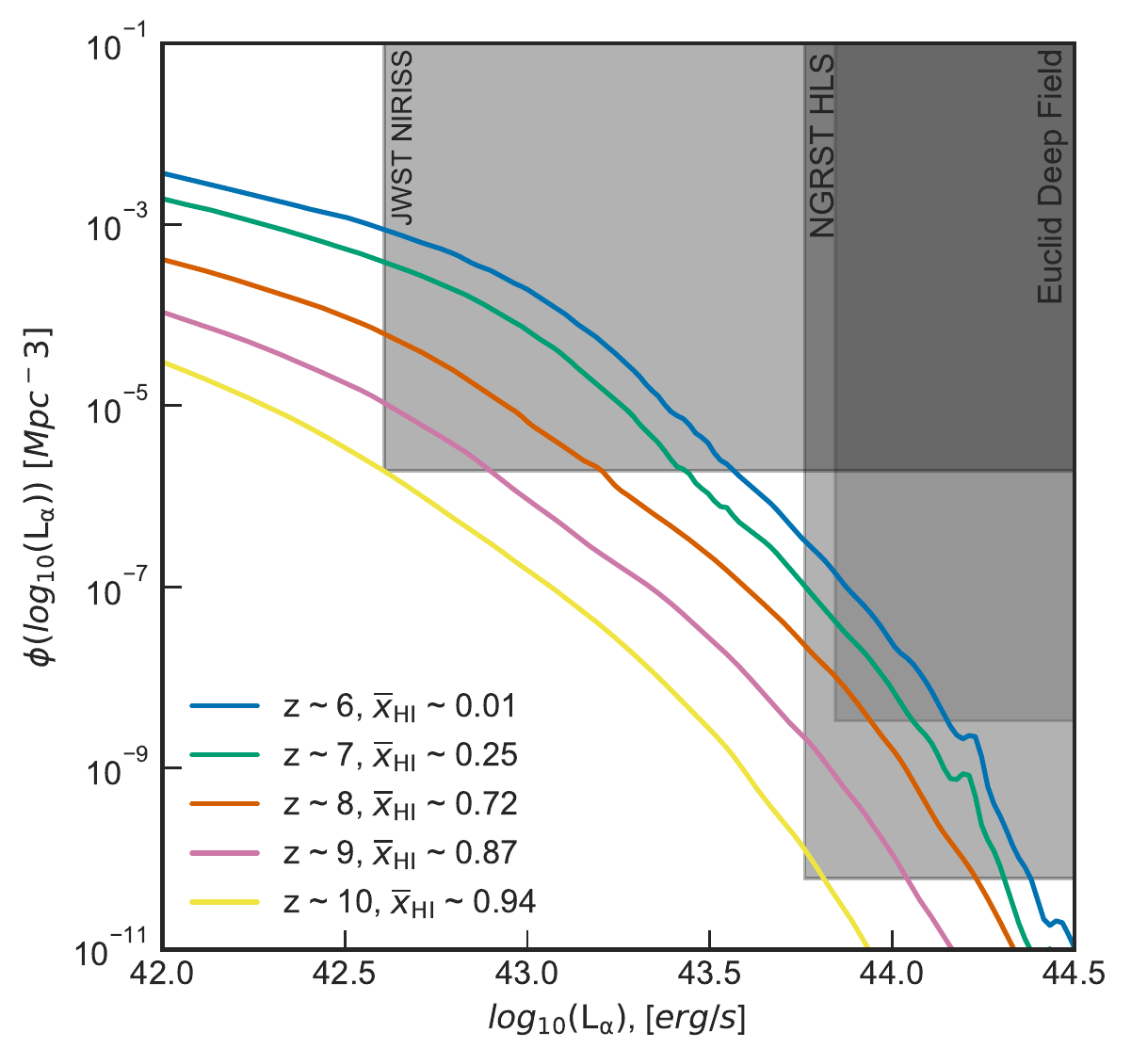}}
    \caption{\lya LF model at redshifts $z = 6 - 10$ and their corresponding median neutral fraction values. The shaded boxes represent the survey depth coverage for the NGRST High Latitude Survey (a $5\sigma$ flux limit of $\sim 7.1\times10^{-17}$ erg s$^{-1}$ cm$^{-2}$ and a redshift range of $8 \leq z \leq 15$ \citep[e.g,][]{Ryan2019,Spergel2013}, the Euclid Deep Field Survey (a $5\sigma$ flux limit of $\sim 8.6\times10^{-17}$ erg s$^{-1}$cm$^{-2}$ and a redshift range of $6 \leq z \leq 10$ \citep[e.g,][]{Laureijs2012,Bagley2017}, and the JWST NIRISS mock pure-parallel survey (a $5\sigma$ flux limit of $\sim 5.0\times10^{-18}$ erg s$^{-1}$ cm$^{-2}$ and a redshift range of $z \gtrsim 5$). We calculate the luminosity limits and depths for the NGRST HLS, Euclid DFS, and JWST NIRISS mock pure-parallel survey at a median redshift value of $z = 8$. }
    \label{fig:lfpred610}
\end{figure}

We make predictions for these surveys in Figure~\ref{fig:lfpred610}. We plot our model \lya LF and the approximate median neutral fraction value based the reionization history allowed by the CMB optical depth and dark pixel fraction at each redshift \citep{Mason2019b} between $ 6 < z < 10$ (shown as the gray shaded region in Figure~\ref{fig:reionization_history}). We see that these surveys will detect luminous \lya emitters at higher redshifts. We predict the NGRST HLS will be able to discover galaxies $L_\alpha > 10^{43.76}$ erg s$^{-1}$ at redshifts up to $z = 10$. Euclid Deep Field survey will be able to detect bright galaxies at $L_\alpha > 10^{43.84}$ erg s$^{-1}$ but only up to $z \sim 8$.
Using the JWST mock pure-parallel survey, we estimate it be able to detect bright galaxies at $L_\alpha > 10^{42.61}$ erg s$^{-1}$ up to $z \sim 9-10$.

We note that the predicted number counts will likely be higher than shown in Figure~\ref{fig:lfpred610} due to gravitational lensing magnification bias, which can increase the observed number of galaxies at the bright-end of the LF in flux-limited surveys \citep{Wyithe2011,Mason2015a,Marchetti2017}. 

\section{Discussion}
\label{sec:disc}
In this Section, we discuss uncertainties that affect our \lya LF model (Section~\ref{disc_caveats}), a comparison with previous work that attempted to constrain reionization from \lya LFs (Section~\ref{disc_compare}), and an explanation of our \lya luminosity distributions for both \lya and UV-selected galaxies and their affect on the normalization factor (Section~\ref{sec:disc_F}). 
\subsection{Modelling caveats} \label{disc_caveats}

In building our model, we make several assumptions that can affect the results which are summarized here. These assumptions are also described by \citet{Mason2018,Whitler2020} and we refer the reader there for additional details. Firstly, we assume that the intrinsic `emitted' \lya luminosity distribution does not evolve with redshift (only \MUV) and is the same as the observed \lya luminosity distribution at $z\sim6$ \citep[as modelled from the EW distribution by][]{Mason2018}, and that only evolution in the `observed' luminosity distribution is due to reionization alone. However, we should expect some evolution in the \lya luminosity distribution with redshift, as galaxy properties evolve. Physically, we could expect galaxies to have higher luminosities as redshift increases, due to, e.g., lower dust attenuation \citep{Hayes2011}, which leads to a steepening of the \lya LF with redshift \citep[][]{Gronke2015,Dressler2015}. In that case, more significant absorption by the IGM would be required to explain the observed \lya LFs and we would thus infer a higher inferred neutral fraction. Alternatively, a decrease in outflow velocities possibly associated with a decreasing specific SFR, could lower \lya escape from galaxies, decreasing the emitted luminosity \citep{Hassan2020}, in which case a lower neutral fraction would be inferred.

We also assume that \lya visibility evolution between $z = 6-7$ is due only to the evolution of the \lya damping wing optical depth \citep[e.g.,][]{Miralda1998}, due to the diffuse neutral IGM. We do not model redshift evolution of the \lya transmission in the ionized IGM or CGM at fixed halo mass \citep[e.g.,][]{Laursen2011,Weinberger2018}. The amount of transmission due to these components is determined by the \lya line shape. If the \lya line velocity offset from systemic decreases with redshift at fixed halo mass, this would decrease the \lya transmission throughout the ionized IGM and CGM \citep[][]{Dijkstra2011,Choudhury2015} and reduce the need for a highly neutral IGM. A full exploration of the degeneracies and systematic uncertainties due to the \lya emission model is left to future work.

\subsection{Comparison with previous work} 
\label{disc_compare}

In this work we directly compare measurements of the \lya LF to our model. Previous works, e.g. \citet{Ouchi2010,Zheng2017,Konno2014,Konno2018,Inoue2018,Hu2019}, estimate the neutral fraction by evaluating the \lya luminosity density. While this provides a reasonable first-order estimate, this method can be difficult to interpret as it relies on models and observations using the same luminosity limit in the luminosity density integral, and it collapses any valuable information that is obtained in the evolving \textit{shape} of the \lya LF (including increasing the statistical uncertainty on \xHI by reducing the number of data points -- see Appendix~\ref{appendix_xHI}). As demonstrated in Section~\ref{sec:results_evo_schech} the \lya shape does evolve.

Furthermore, these works estimate the neutral fraction based on the assumption that the transmission fraction, $T_{\mathrm{IGM}}$ is constant for all galaxies. However, as shown by e.g. \citet{Mason2018,Whitler2020} it is a broad distribution and depends on galaxy properties, through their large scale structure environment and the internal kinematics that sets the \lya line shape. If the transmission fraction varies with \lya or UV luminosity, the common method of calculating \lya transmission by taking the ratio of the \lya luminosity density at different redshifts is invalid, thus our work enables a more robust estimate of \xHI from the \lya LFs. Many of these works compare with simulations by \citet{McQuinn2007}, which do model the impact of inhomogeneous reionization on the \lya LF but take a more simplistic approach to modelling \lya luminosity: each galaxy has a \lya luminosity proportional to its mass. In our work, we use the EW probability distribution from \citet{Mason2015}, to take into account that galaxies have a range of \lya EW at fixed UV magnitude. Our approach is thus more similar to that of \citet{Jensen2013} who model the \lya luminosity probability distribution as a function of halo mass. However, \citet{Jensen2013} model the \lya luminosity and equivalent width distributions independently, whereas we have shown the \lya LF can be self-consistently described by the same \lya EW distribution that describes Lyman-break galaxies.

Finally, our semi-analytic model provides flexibility over approaches which model radiative transfer in N-body simulations \citep[e.g.,][]{Dayal2011,Jensen2013,Hutter2014,Inoue2018} or sophisticated hydrodynamical simulations \citep[e.g.,][]{Dayal2011,Weinberger2019}, by keeping the IGM neutral fraction and redshift as free parameters, rather than assuming a fixed reionization history.
Using this new model for the \lya LF, we can separate redshift and the neutral fraction, fixing either parameter if needed, and see how observations compare to our model.

\subsection{Reconciling the \lya luminosity distributions for \lya and UV-selected galaxies}
\label{sec:disc_F}

As described in Section~\ref{sec:mod_normLF}, a normalization factor, $F$, is introduced to the \lya LF to account for any mismatch in the number density of LAEs \citep{Dijkstra2012}. If the \lya luminosity distribution model accurately describes the Lyman-break galaxy population observed in the UV LF, we expect $F=1$. We find $F=0.974$ in our model, which is considerably higher than previous work focusing on the \lya LF at lower redshifts which found $F\sim0.5$ \citep{Dijkstra2012,Gronke2015}.

The key difference compared to this previous work which leads to our model successfully reproducing the \lya LF without the need for additional normalization, is due to the EW distribution we employed.
While we, as well as the previous work, include non-emitters in the model, the EW distribution used by \citet{Dijkstra2012,Gronke2015} \citep[calibrated to measurements of][at $z\sim 3-6$]{Shapley2003,Stark2010,Stark2011} shows an increasing probability of \lya emission for lower UV brightness galaxies, leading to a \lya emitter ($EW > 0$\,\AA) fraction of unity for $\MUV \simgt -19$. The EW distribution we used caps this `emitter fraction' at $65\%$ (our function $A(\MUV)$ in Section~\ref{sec:mod_lumpd}). 

Which parametrization of the EW distribution is most appropriate is still up to debate \citep[see, e.g., detailed discussion by][who study more complex distributions also dependent on the stellar mass and the UV slope]{Oyarzun2017}. The recent study of \citet{Caruana2018} supports our non-emitter fraction, as they find a fraction of $\sim 0.5\pm 0.15$ galaxies with $EW > 0\,$\AA\ at $3<z<6$ in HST continuum selected galaxies for within the MUSE Wide field \citep{Herenz2017,Urrutia2019} is present. However, \citet{Caruana2018} also find a non-evolution of this fraction with UV magnitude (in contrast with previous models) as well as typically lower \lya fractions for larger EW cuts (e.g., the value for $W > 50\,$\AA\ seems to be in slight tension with measurements by \citealp{Stark2010} who find $\approx 45\pm 10\%$ of galaxies have \lya $EW > 55\,$\AA\ and a strong anti-correlation with UV brightness).

Another factor to consider when comparing $z\lesssim 3$ and $z\gtrsim 5$ EW distributions is the impact of the IGM on the \lya line even at $z\sim 5-6$. In fact, is has been suggested by \citet{Weinberger2019} that $F$ could stem from this effect but note that \citet{Dijkstra2012} and \citet{Gronke2015} compare their modelled \lya LFs at $z\sim 3$ to data by \citet{Ouchi2008,Gronwall2007} and \citet{Rauch2008}, respectively, and still require $F\sim 0.5$ at this low $z$.

Future observational studies will constrain the \lya EW distribution further both as a function of redshift and UV magnitude, and thus can quantify the fraction of non-emitters for UV faint galaxies. We have shown that a constant non-emitter fraction of $\sim 35\%$ for $M_{\rm UV}\lesssim -19$ makes the fudge factor $F\sim 0.5$ obsolete, which indicates that such a `cutoff' exists in reality.

\section{Conclusions}\label{sec:conc}

We have developed a model for the \lya LF during reionization, and compared it with observations at specific redshifts to estimate the evolution of the neutral fraction. Our model can be extended to predict the evolution of the \lya LF with neutral fraction at even higher redshifts, deeper in the era of reionization. Our model takes into account inhomogeneous reionization, enabling us to understand the impact of galaxy environment on the \lya LF.

Our conclusions are as follows:
\begin{enumerate}
    \item By combining previously established models for the UV luminosity function and the \lya EW distribution for UV-selected galaxies, we successfully reproduce the observed $z=5.7$ \lya luminosity function (derived from \lya-selected galaxies).
    
    \item Our model predicts a decline in the \lya luminosity function as the neutral fraction increases. For $\xHI \simlt 0.4$, the \lya LF models exhibit relatively little decrease in number density, however, at higher neutral fractions we see a significant drop in number density.

    \item We predict that the average \lya luminosity for a Lyman-break galaxy of a given UV magnitude decreases as the neutral fraction increases. We find there is only a moderate decrease in \lya luminosity for UV-bright galaxies at increasing \xHI (factor of $\sim 2$ from a fully ionized to fully neutral IGM), because they typically exist in dense regions of the universe that reionize early, allowing large amounts of \lya photons to be transmitted. For UV-faint galaxies which are typically found in neutral IGM regions, we see a decrease in \lya luminosity by a factor of $\sim10$ with the neutral fraction.    
    
    \item We find strong evolution of the Schechter function parameters with \xHI, demonstrating the LF shape changes. The faint-end slope $\alpha$, number density $\phi^*$ and the characteristic luminosity $L^*$ all generally decrease with increasing neutral fraction. These decreases in the Schechter function parameters with increasing \xHI can be explained by a reduction in \lya luminosity from all galaxies, with the faintest galaxies experiencing the most significant decline in transmission which shifts the faint-end slope to steeper values.

    \item The \lya luminosity density decreases overall as the universe becomes more neutral, as shown by previous work.
    
    \item We perform a Bayesian inference of the IGM neutral fraction given observations using our model. We infer an IGM neutral fraction at $z=6.6$ of $\xHI= 0.08^{+0.08}_{-0.05}$, rising to $\xHI= 0.28 \pm 0.05$ for $z = 7.0$ and $\xHI=0.83^{+ 0.06}_{- 0.07}$ for $z = 7.3$, providing further evidence for a late and fairly rapid reionization.
    
    \item Using our \lya LF model with a fiducial reionization history, we predict the NGRST HLS will be able to discover bright galaxies with $L_{\alpha} > 10^{43.76}$ erg s$^{-1}$ at redshifts up to $z = 10$. Euclid Deep Field survey will be able to detect bright galaxies at $L_{\alpha} > 10^{43.84}$ erg s$^{-1}$ but only up to $z \sim 8$. Using a JWST mock pure-parallel survey, we estimate it be able to detect galaxies at $L_\alpha > 10^{42.61}$ erg s$^{-1}$ up to $z \sim 9-10$.
\end{enumerate}

Constraining the evolving shape of the \lya LF as a function of redshift provides an important tool to estimate the evolving neutral fraction during reionization. Understanding the timeline of reionization and the properties of galaxies that existed as a function of redshift, and how they are impacted by neutral gas, can ultimately be used to infer properties of the first stars and galaxies that initialized reionization.

\acknowledgements

The authors thank Masami Ouchi, Weida Hu, Kazuaki Ota, Akio Inoue, Yoshiaki Ono, and Takatoshi Shibuya for sharing the \lya LF observations from \citet{Ouchi2008,Ouchi2010, Hu2019, Ota2017, Itoh2018, Shibuya2012}. AMM thanks Matthew Ashby and Jonathan McDowell for their support throughout the SAO REU program and their feedback when revising early drafts of this paper. 

The SAO REU program is funded in part by the National Science Foundation REU and Department of Defense ASSURE programs under NSF Grant no.\ AST-1852268, and by the Smithsonian Institution. CAM acknowledges support by NASA Headquarters through the NASA Hubble Fellowship grant HST-HF2-51413.001-A awarded by the Space Telescope Science Institute, which is operated by the Association of Universities for Research in Astronomy, Inc., for NASA, under contract NAS5-26555. CS and SB acknowledge the support from Jet Propulsion Laboratory under the grant award RSA 1646027. MG was supported by NASA through the NASA Hubble Fellowship grant HST-HF2-51409 and acknowledges support from HST grants HST-GO-15643.017-A, HST-AR15039.003-A, and XSEDE grant TG-AST180036.

\noindent
\textit{Software}: \verb|IPython| \citep{Perez2007a}, \verb|matplotlib| \citep{Hunter2007a}, \verb|NumPy| \citep{VanderWalt2011a}, \verb|SciPy| \citep{Oliphant2007a}, \verb|Astropy| \citep{Robitaille2013}, \verb|emcee| \citep{Emcee}.

\bibliographystyle{mnras}
\bibliography{library} 

\appendix 
\label{appendix}
\section{Example of fitting Schechter function parameters}\label{appendix_schechter}

\begin{figure*}[h]
    \centering
    \includegraphics[scale=0.35]{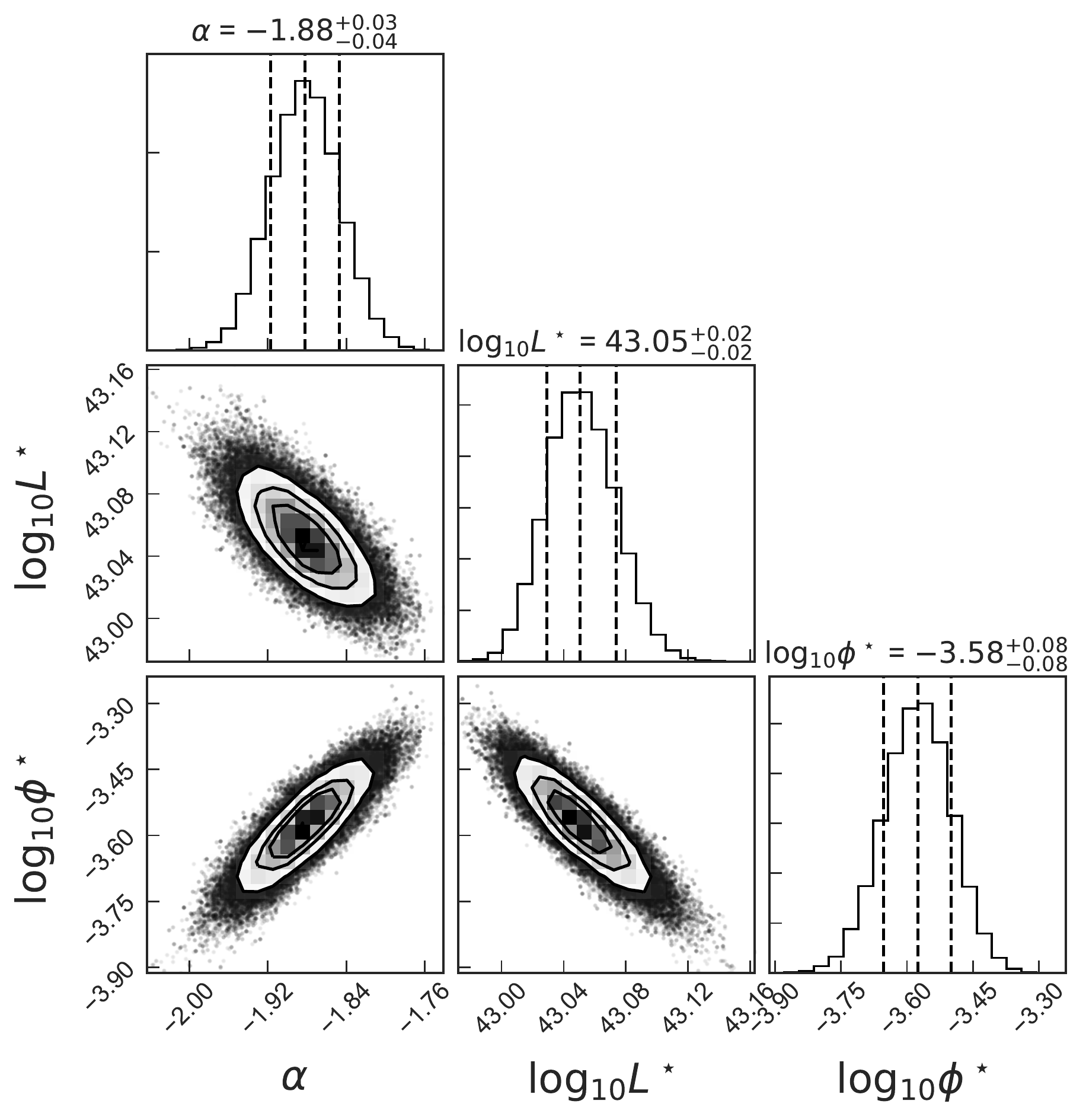}
    \includegraphics[scale=0.6]{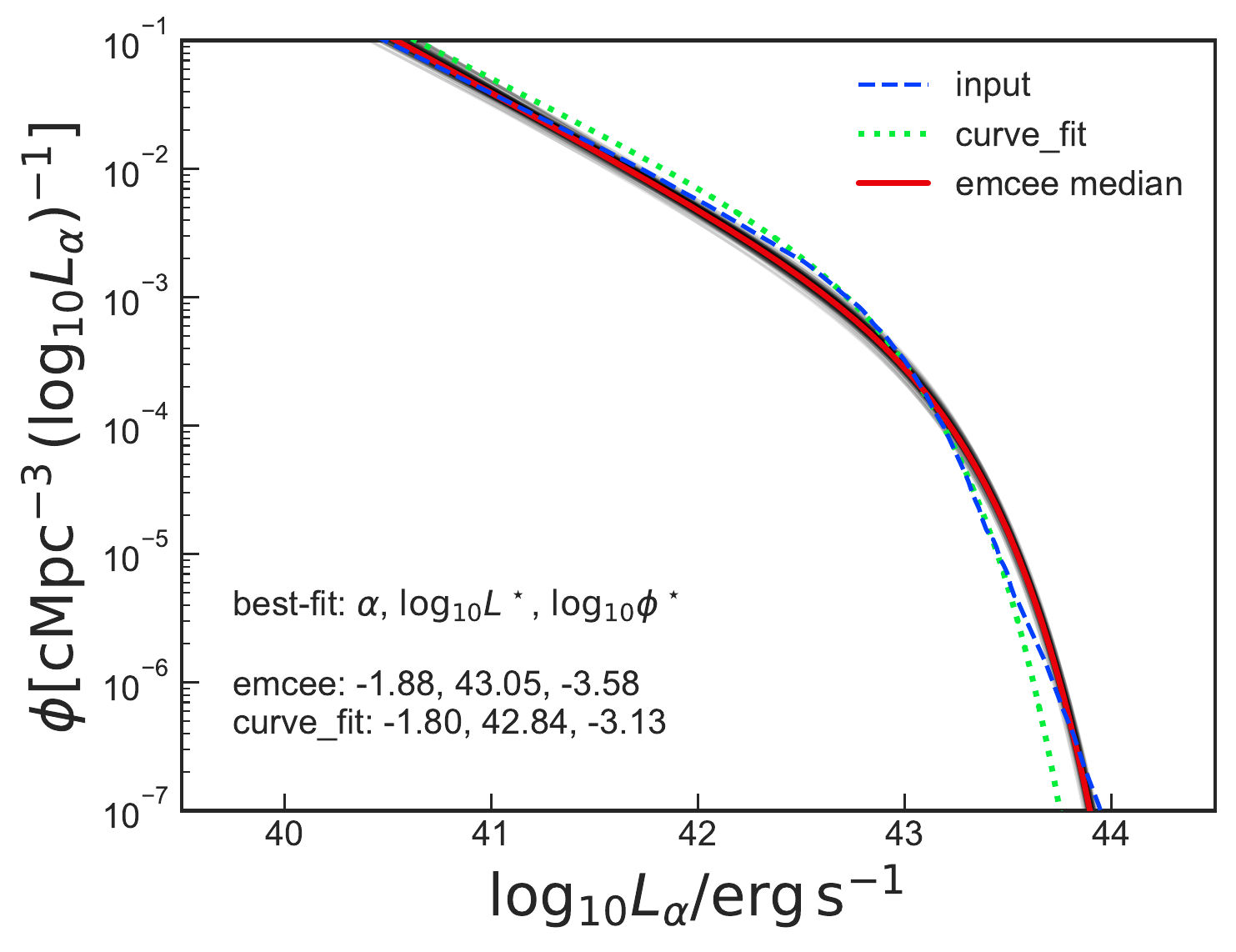}
    \caption{(Left) Posterior probability distributions for Schechter function parameters fit to our \lya LF model at $z = 5.7$ and $\xHI = 0.01$. (Right) Our \lya LF model at $z = 5.7$ and $\xHI = 0.01$ (blue dashed line) compared with the best-fit Schechter functions. Thin black solid lines show Schechter LFs with parameters from 100 draws from the posterior distribution. The Schechter LF obtained from the median of the posteriors is shown as the red solid line. We also plot the Schechter fit obtained using \texttt{SciPy curve\_fit} for comparison (green dotted line). We use the median values of the Schechter function parameters for our analysis.}
    \label{fig:Emceeplots}
\end{figure*}

Figure~\ref{fig:Emceeplots} shows examples of outputs for the best-fit Schechter function parameters for $z = 5.7$ and $\xHI = 0.01$ as described in Section~\ref{sec:results_evo_schech}. We perform a Bayesian inference (Equation~\ref{eqn:bayes}) to obtain the parameters, using the likelihood:
\begin{equation} \label{eqn:schechter_likelihood}
P(\phi \,|\, \alpha, L_\star, \phi_\star) \propto \prod_i \exp{\left[-\left(\ln{\phi_\mathrm{mod}(L_i)} - \ln{\phi_\mathrm{Sch}(L_i, \alpha, L_\star, \phi_\star)} \right)^2 \right]}
\end{equation}
where $\phi_\mathrm{mod}(L_i)$ are our model \lya LFs at luminosity values $L_i$, and $\phi_\mathrm{Sch}(L_i, \alpha, L_\star, \phi_\star)$ is the Schechter function (Equation~\ref{eqn:schechter_LF}). We perform the inference using \verb|emcee| \citep{Emcee}.

For all Schechter function fits, we restrict the fit to the luminosity range $41.0 < \log_{10}L_\alpha < 44.0 $ erg s$^{-1}$, and use the following uniform priors for the Bayesian inference: $-4 < \alpha < 0$, $ 40 < \log_{10}L_\star < 44$ and $-10 < \log_{10}\phi_\star < -2$. Example posteriors and fitted LFs are shown in Figure~\ref{fig:Emceeplots}. We explored a variety of luminosity ranges for the fitting and note that the absolute values of the recovered Schechter parameters are quite sensitive to the fitted luminosity range, however the trends in redshift and neutral fraction are consistent across the luminosity ranges, as long as $<L_\star$ luminosities were included. Our model LFs have $L_\star \sim 10^{43}$\,erg s$^{-1}$, comparable to the luminosity limits of $z\simgt7$ surveys \citep[e.g.,][]{Ota2017,Hu2019}, demonstrating the importance of deep LAE surveys to obtain accurate fits to the observed luminosity functions. In our analysis in Section~\ref{sec:results_evo_schech} we use the use the median values of the Schechter function parameters obtained from the posteriors. Note that our model for the \lya LF is not well described by a Schechter fit -- we see a shallower bright-end drop off.

\section{Inference of the neutral fraction, \xHI at $z = 6.6, 7.0, 7.3$}\label{appendix_xHI}

\begin{figure}[htb]
  \includegraphics[width=.45\linewidth]{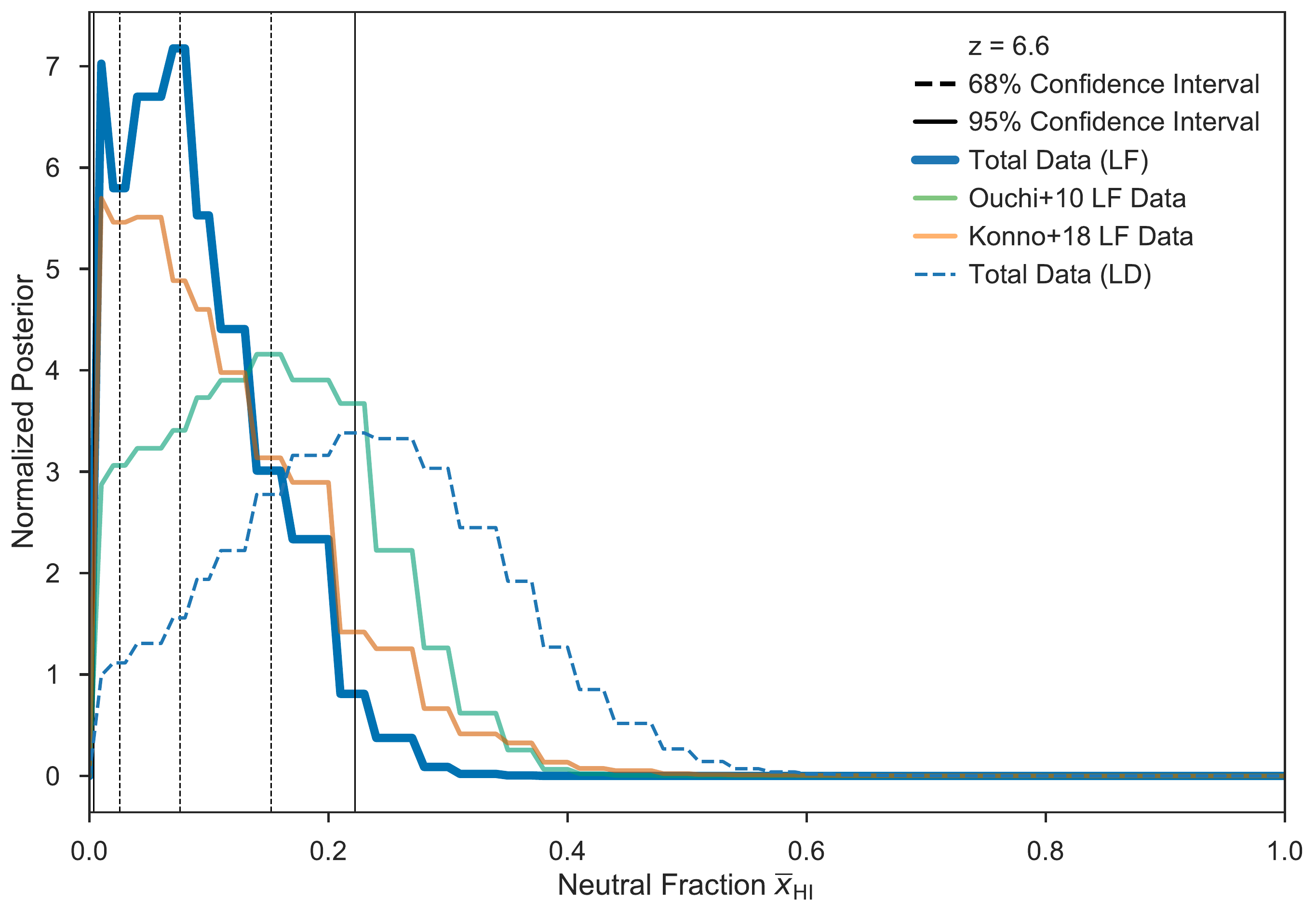}%
\hfill 
  \includegraphics[width=.475\linewidth]{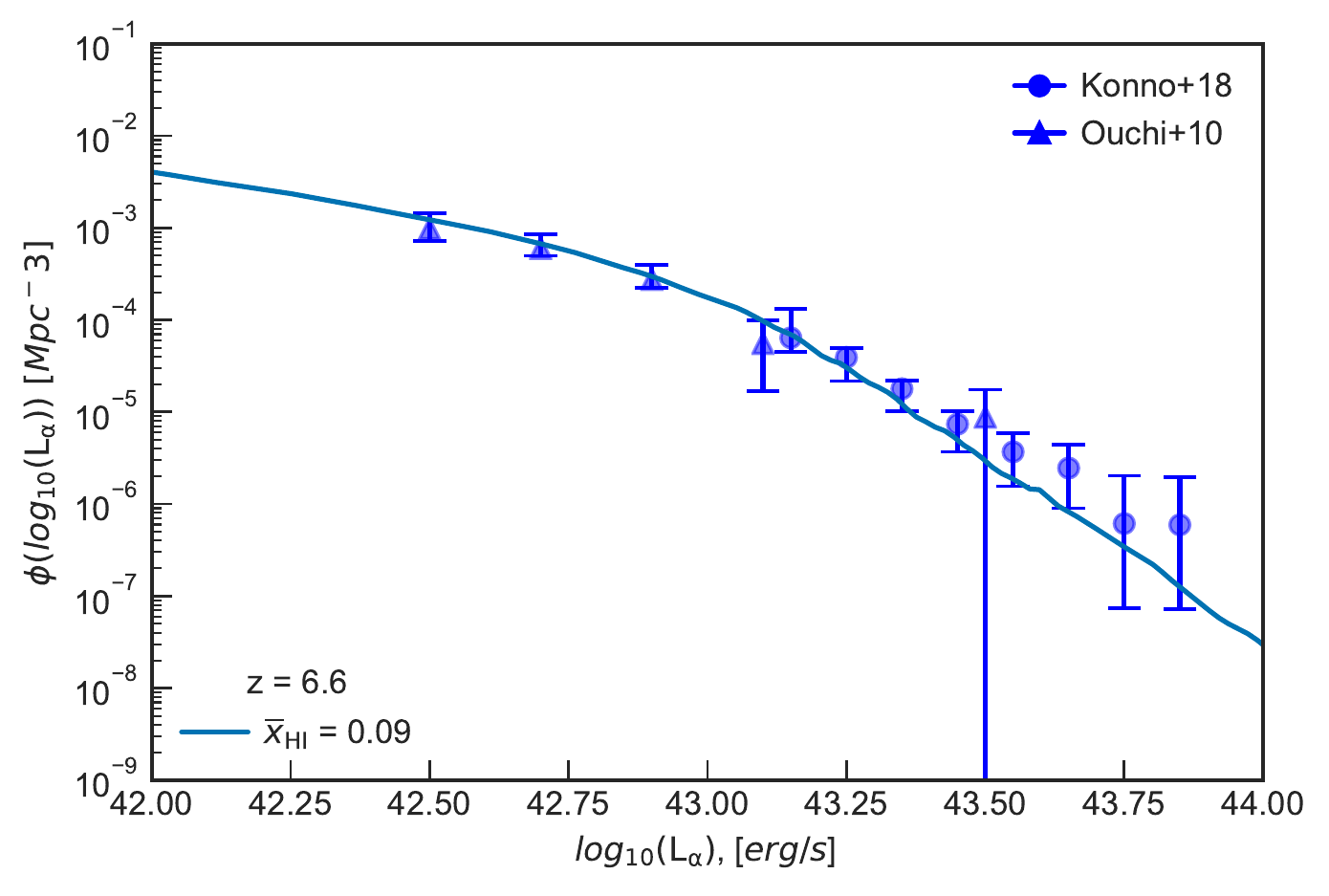}
  \includegraphics[width=.45\linewidth]{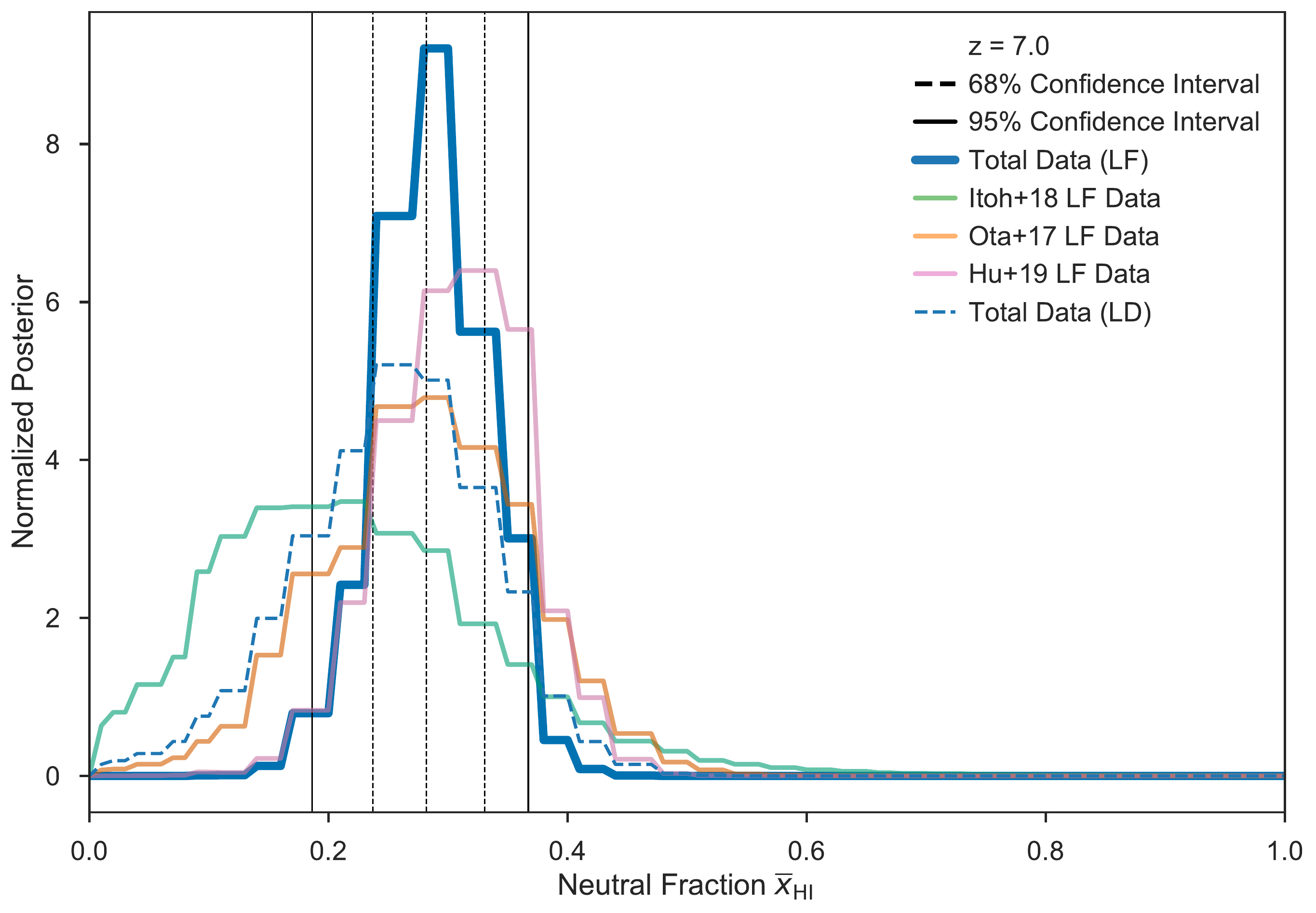} %
\hfill
  \includegraphics[width=.475\linewidth]{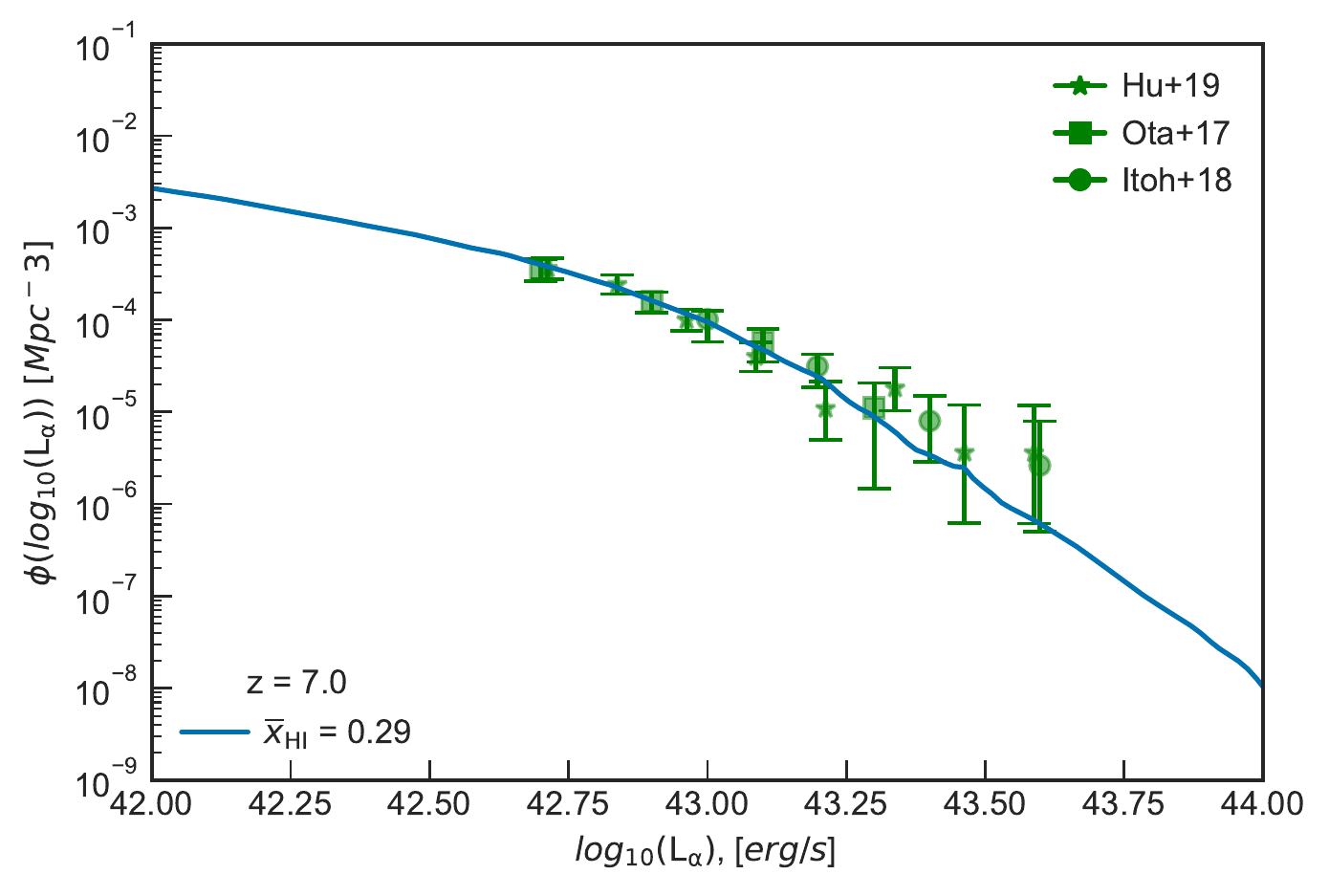}
  \includegraphics[width=.45\linewidth]{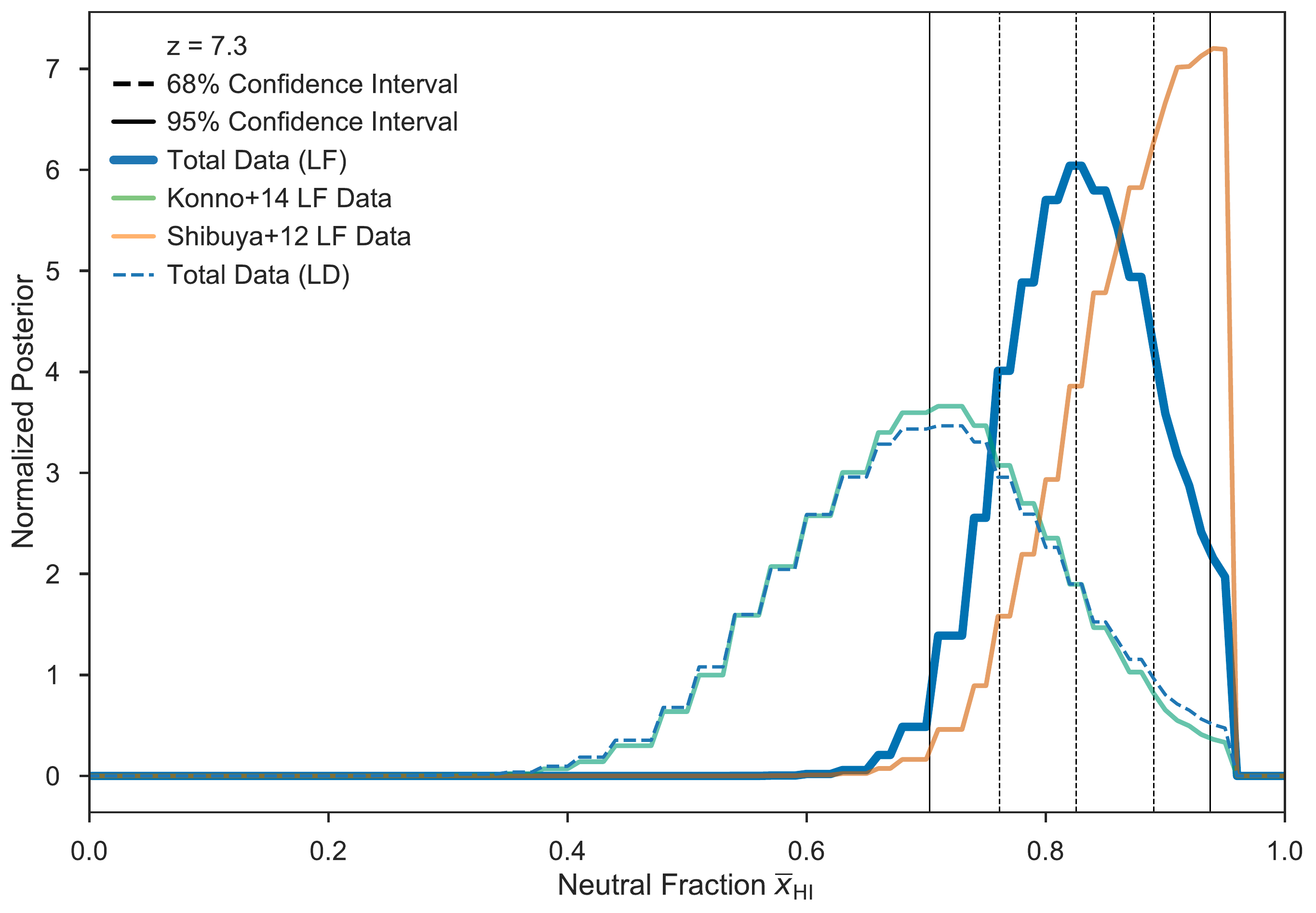}%
\hfill
  \includegraphics[width=.475\linewidth]{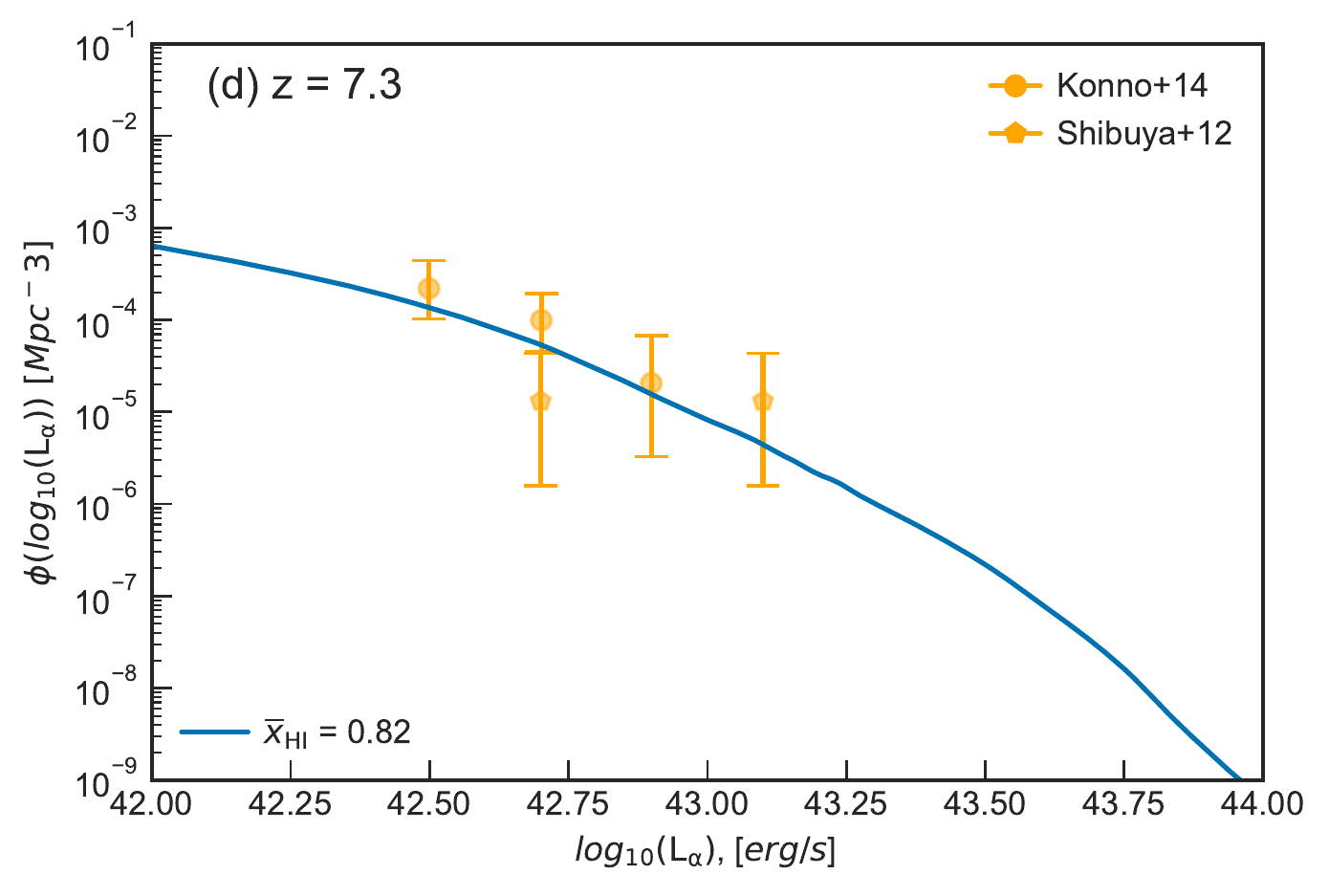}  
\caption{(Left) Neutral hydrogen fraction versus the normalized posterior distribution $P(\xHI|\phi_\mathrm{obs},z)$ (see Section~\ref{sec:mod_bayes}) for each redshift. The thick blue lines in each posterior plot show the total posterior distributions at each redshift alongside the individual posterior distributions for each data set (shown in either green, orange, or pink thin lines). The blue dashed lines correspond to total \lya LD posterior distributions. The black dashed lines show the $68 \%$ or $1 \sigma$ confidence interval where the upper, lower, and median limits are defined. The black solid lines show the upper and lower limits for the $95 \%$ or $2 \sigma$ confidence interval. For redshifts $z = 6.6, 7.0, 7.3$, our median neutral fraction values (mid-dashed line on the left figures) is estimated to be: $\xHI(z = 6.6) = 0.08 $, $\xHI(z = 7.0) = 0.28$, and $\xHI(z = 7.3) = 0.83$. If we compare our total data posteriors, the \lya LD posterior distribution has median neutral fraction values of: $\xHI(z = 6.6) = 0.22 $, $\xHI(z = 7.0) = 0.25$, and $\xHI(z = 7.3) = 0.69$. In Figure~\ref{fig:reionization_history} we plot the median and $68 \%$ confidence interval values. (Right) We show the corresponding \lya LF plot at that redshift and median neutral fraction value to verify our fit.}
    \label{fig:xHIminimization}
\end{figure}

To infer the neutral fraction we calculate a posterior distribution for $z = 6.6, 7.0, 7.3$.  The posterior, defined in Section~\ref{sec:mod_bayes}, is normalized and plotted against neutral fraction values $\xHI$. To determine the $68 \%$ and $95 \%$ confidence intervals, we interpolate the inverse cumulative distribution function (CDF) to find the neutral fraction value that falls at a given confidence interval value. Figure~\ref{fig:xHIminimization} shows the \xHI posterior at each redshift for both \lya LF data and \lya LD data, along with the confidence intervals. We also compare the \lya LF models at the median \xHI values to the data to verify our inferred values.

To verify advantages the \lya LF may have over the \lya LD in estimating the median neutral fraction, we establish the \xHI posterior using the \lya LD data. We expect larger uncertainties in the neutral fraction using only the LD data -- the errors should roughly increase by $\sqrt{N}$, where $N$ is the ratio of the number of individual data points with the two methods. Also note, some observations do not have \lya LD data that can be included in the estimation of the neutral fraction \citep[e.g,][]{Ouchi2010, Shibuya2012} and thus also affect the results.

\end{document}